\documentclass[sigconf]{acmart} 
\usepackage{float}
\usepackage{stfloats}

\AtBeginDocument{%
  \providecommand\BibTeX{{%
    \normalfont B\kern-0.5em{\scshape i\kern-0.25em b}\kern-0.8em\TeX}}}

\copyrightyear{2024}
\acmYear{2024}
\setcopyright{rightsretained}
\acmConference[CHI '24]{Proceedings of the CHI Conference on Human Factors in Computing Systems}{May 11--16, 2024}{Honolulu, HI, USA}
\acmBooktitle{Proceedings of the CHI Conference on Human Factors in Computing Systems (CHI '24), May 11--16, 2024, Honolulu, HI, USA}
\acmDOI{10.1145/3613904.3642608}
\acmISBN{979-8-4007-0330-0/24/05}

\begin{document}

\title[Making Sense of Waiting Time Activities]{Fragmented Moments, Balanced Choices: How Do People Make Use of Their Waiting Time?}

\author{Jian Zheng}
\affiliation{%
  \institution{University of Maryland}
  %\streetaddress{P.O. Box 1212}
  \city{College Park}
  \state{Maryland}
  \country{USA}
  \postcode{20742}
}
\email{jzheng23@umd.edu}
\orcid{0000-0002-6075-9313}

\author{Ge Gao}
\affiliation{%
  \institution{University of Maryland}
  %\streetaddress{P.O. Box 1212}
  \city{College Park}
  \state{Maryland}
  \country{USA}
  \postcode{20742}
}
\email{gegao@umd.edu}
\orcid{0000-0003-2733-2681}

%\renewcommand{\shortauthors}{Zheng and Gao}

\begin{abstract}
  Everyone spends some time waiting every day. HCI research has developed tools for boosting productivity while waiting. However, little is known about how people naturally spend their waiting time. We conducted an experience sampling study with 21 working adults who used a mobile app to report their daily waiting time activities over two weeks. The aim of this study is to understand the activities people do while waiting and the effect of situational factors. We found that participants spent about 60\% of their waiting time on leisure activities, 20\% on productive activities, and 20\% on maintenance activities. These choices are sensitive to situational factors, including accessible device, location, and certain routines of the day. Our study complements previous ones by demonstrating that people purpose waiting time for various goals beyond productivity and to maintain work-life balance. Our findings shed light on future empirical research and system design for time management.
\end{abstract}





\maketitle

\section{Introduction}
Every day, everyone experiences moments of waiting, whether for an appointment, a meeting, waiting for others, or standing in line at a store. Waiting requires delaying action until a particular time or event occur and it can happen almost any time, any place. Typically individuals lack specific pre-assigned tasks during these waiting periods, so they are relatively free to choose whatever task they deem appropriate.

Two threads of research in human-computer interaction (HCI) have delved into waiting time usage. Studies on wait-learning have developed mobile phone and computer applications enabling users to learn vocabulary while waiting, for example, for an elevator or a message \cite{Cai2015-sy,Cai2017-gc}. In parallel, microtask research, while not exclusively targeting waiting time, has explored opportunities to break complex tasks down into smaller sub-tasks \cite{August2020-ew} and accomplish them in micromoments, or short bursts of time traditionally seen as unproductive \cite{Williams2019-ls}.  
However, there has been little empirical knowledge about how people naturally spend their waiting time: What activities do they spontaneously engage in and how do situational (e.g., time and location) factors impact these activities? Without knowing how and why people spend their waiting time, it is impossible to assess the effectiveness of developed tools in improving the use of that time. This knowledge gap hinders researchers from evaluating intervention effectiveness and tailoring strategies to diverse situations. To fill in this gap, we aim to answer two research questions (RQ) through this current study:

\begin{itemize}
\item RQ1: What do people do while waiting?
\item RQ2: How do situational factors affect people's waiting time activities?
\end{itemize}

We conducted an experience sampling method (ESM) study to address the research questions. Twenty-one participants were recruited to report their waiting time activities at least three times a day, either when they found themselves waiting for something or when they received a prompt, for two weeks. Participants reported situational factors including waiting duration, location, time of the day, and devices available to them during waiting periods. Our findings indicate that people allocate approximately 60\% of their waiting time to leisure activities (e.g., games, videos, and music), 20\% to productive activities (e.g., work and study), and 20\% to maintenance activities (e.g., housework and personal care). When waiting at workplaces and/or with computer access, people are more likely to do productive activities; when waiting at home and/or without mobile phone access, people are more likely to do maintenance activities. During lunch breaks, people are less likely to do leisure activities while waiting.
Data obtained from our study demonstrated the distribution of waiting time activities in a natural setting and the effects of situational factors. Findings of our work shed light on the design of time management tools and protocols that can adapt to individual needs as well as situational factors. They also provide the benchmark for future HCI work that evaluates people’s waiting time activities and how they vary in response to technical interventions.

\section{Related Work}
\subsection{Time Usage Studies}
A long history of research in HCI and beyond has been looking into time usage \cite{Lindley2015-hv, Odom2018-dv, Rahm-Skageby2022-iv, Wiberg2021-tx}. For instance, Lindley \cite{Lindley2015-hv} illustrated how digital technologies influence our perception of time and how the fragmented routines of modern Western societies shape this experience. Wiberg and Stolterman \cite{Wiberg2021-tx}, upon a literature review, identified what aspects about time and temporality have been studied and how they have been studied in the HCI field. Broadly, the American Time Use Survey (ATUS \cite{ATUS2022-bc}) provides nationally representative data on how United States residents aged 15 and above spend their time. Its sample are "drawn from the population of households that participated in the Current Population Survey" \cite{ATUS2021-mw}. Similarly, the Multinational Time Use Study (MTUS \cite{MTUS2023-os}) aggregate data from over a million diary days from more than 100 nationally representative surveys, offering insights into how individuals from diverse countries allocate their time. Other work has investigated time allocation patterns within specific demographics, such as adolescents \cite{Csikszentmihalyi1984-kj}. 

Across these studies, researchers do not typically directly analyze specific activities reported by participants. Instead, they categorize activities into three primary domains: productive, maintenance, and leisure activities, each with several sub-categories. Productive activities are those related to paid work or schoolwork; most studies’ sub-categories include "work" and "study". Productive activities are done for direct payment or self-growth. Maintenance activities are those required to keep ones daily life functioning acceptably \cite{Csikszentmihalyi1984-kj}, such as sleeping and having meals; "housework" and "personal care" are common sub-categories. Leisure activities are those individuals engage in for enjoyment or well-being \cite{Su2022-nm}. Whereas productive and maintenance activities are duties and responsibilities, leisure activities are neither.

\subsection{Gap Time Between Scheduled Events}
People’s daily awake time can be roughly divided into scheduled events and the unscheduled gaps between. During scheduled events, people are often required to perform predefined tasks or duties. For example, during scheduled workplace meetings, people engage in work communication and pay peripheral attention to other tasks \cite{Cao2021-ao}. During gym time, people are supposed to be working out. In contrast, usage of gap time is more personal and subject to an individual’s needs at that time \cite{Rattenbury2008-gz}. Activities undertaken during these gaps can reflect diverse individual needs, motivations, and even personalities.

The above findings lead HCI scholars to the question of how gap time can be best managed to satisfy a person’s situational needs. A growing number of recent studies have examined the use of gap time in some specific formats. During work breaks, for instance, knowledge workers have been found to leverage information and communications technology to fulfil home responsibilities from their offices, aiming to strike a balance between work and life \cite{Skatova2016-lx}. In the case of public transit time, people adapt their tasks to fit the commuting environment so that they can effectively engage in routine activities and perform light work \cite{Lee2023-xo}. 
 
\subsection{Waiting Time as an Underexplored yet Ubiquitous Format of Gap Time}
One underexplored format of gap time is waiting time, different from other gap time in multiple aspects. First, waiting is ubiquitous. Different from work breaks and transit time, waiting can take place at any moment with or without foresight and is not tied to any specific location. Second, waiting can often be “plastic” \cite{Rattenbury2008-gz}. That is, the duration of some wait times are less predictable than others (e.g., waiting in an emergency room vs. waiting to see a doctor at an appointment). As a result, people need to make on-the-fly decisions about the activities to perform during the waiting time.

Previous research has developed tools and systems that enable people to leverage their waiting time for productivity. For instance, WaitChatter \cite{Cai2015-sy} displays contextually relevant foreign language vocabulary and micro-quizzes while users wait for message responses. Field testing showed that  users learned an average of 57 new words in two weeks. WaitSuite \cite{Cai2017-gc} incorporated WaitChatter along with four other wait-learning apps, allowing users to learn vocabulary during various waiting scenarios, such as elevator rides, mobile app content loading, establishment of a WiFi connection, and email sending delays. Zaturi \cite{Kang2017-cf} enables parents to create audiobooks for their babies in their gap times. 

Another relevant line of research discusses waiting time usage from the angle of micromoments—particularly small time gaps before the next scheduled event to star \cite{Teevan2016-ed}—and small tasks that can be completed during those moments (microtasking). Previous research in this space examines ways to write a complete document in scattered micromoments \cite{August2020-ew, Hahn2019-zy, Iqbal2018-tj, Swearngin2021-ys}. For example, Play Write \cite{Iqbal2018-tj} allows users to create writing microtasks, such as correcting spelling, identifying wordy sentences, shortening sentences, and accepting or rejecting changes, to be done at gap time between other tasks. In a follow-up study, Play Write was encapsulated into a Chrome extension which embeds these microtasks between posts in a user’s Facebook feed \cite{Hahn2019-zy}. A similar time management mechanism has also been applied to programming \cite{Williams2019-ls}. 

The bulk of above research often assumes that people want to purpose their waiting time  productively. It raises the question of how people spend their waiting time in natural settings as opposed to testing sessions for the evaluation of tools. Specifically, will individuals choose to engage in productive or non-productive activities? Previous research does not answer this question. In this paper, we present our longitudinal research that documents people’s organic waiting time activities in real-world settings and investigates potential influences from situational factors.

\section{Method}

\subsection{Participants}

We recruited participants through Prolific, Facebook groups, and snowball sampling. Qualified participants were (a) located in the United States, (b) 18 years old or above, (c) native English speakers, (d) employed in a full-time, part-time, or freelance job, and (e) users of Android phones that our ESM app is designed for.

Among an initial pool of 39 participants, 16 individuals stopped responding through the ESM app soon after the data collection began; 2 individuals were identified as malicious participants who provided fake profile information (e.g., they were located out of the United States, as indicated by IP address). We therefore excluded these individuals from our data analysis and fully removed their data from the dataset. Table \ref{table:participants} provides detailed information of the remaining 21 participants. This sample size is in line with the local standards of ESM-based research in HCI \cite{van-Berkel2018-wr}. Each participant was compensated with 70 to 100 USD depending on the number of ESM reports they had submitted over the period of the study. 

\begin{table*}
  \caption{Participants' Demographic Information and NO. of ESM Reports}
  \label{table:participants}
    \begin{tabular}{ccccccc}
    \toprule
    ID  & Gender & Age & Employment & Work time       & Work place & ESM reports \\
    \midrule
    P01 & Male      & 28  & Full-time  & 9 a.m. to 5 p.m.          & Hybrid     & 34          \\
    P02 & Male      & 36  & Full-time  & 9 a.m. to 5 p.m.          & Hybrid     & 33          \\
    P03 & Female      & 48  & Full-time  & Rotating shifts & Hybrid     & 38          \\
    P04 & Female      & 33  & Full-time  & 9 a.m. to 5 p.m.          & On-site    & 42          \\
    P05 & Female      & 41  & Full-time  & 9 a.m. to 5 p.m.          & Remote     & 42          \\
    P06 & Female      & 42  & Part-time  & As required     & Hybrid     & 29          \\
    P07 & Male      & 32  & Full-time  & Rotating shifts & On-site    & 43          \\
    P08 & Female      & 46  & Full-time  & 9 a.m. to 5 p.m.          & On-site    & 42          \\
    P09 & Female      & 32  & Part-time  & Rotating shifts & Remote     & 41          \\
    P10 & Female      & 43  & Full-time  & 9 a.m. to 5 p.m.          & On-site    & 32          \\
    P11 & Female      & 50  & Freelance        & As required     & Remote     & 45          \\
    P12 & Female      & 34  & Full-time  & 9 a.m. to 5 p.m.          & On-site    & 47          \\
    P13 & Female      & 31  & Full-time  & Rotating shifts & On-site    & 45          \\
    P14 & Male      & 32  & Full-time  & 9 a.m. to 5 p.m.          & On-site    & 50          \\
    P15 & Male      & 25  & Full-time  & Rotating shifts & Hybrid     & 43          \\
    P16 & Male      & 34  & Freelance        & Rotating shifts & On-site    & 43          \\
    P17 & Female      & 55  & Full-time  & 9 a.m. to 5 p.m.          & Hybrid     & 75          \\
    P18 & Female      & 60  & Full-time  & 9 a.m. to 5 p.m.          & Remote     & 50          \\
    P19 & Female      & 61  & Full-time  & As required     & Hybrid     & 31          \\
    P20 & Female      & 53  & Full-time  & 9 a.m. to 5 p.m.          & On-site    & 25          \\
    P21 & Male      & 52  & Full-time  & 9 a.m. to 5 p.m.          & Remote     & 42          \\
    \bottomrule
    \end{tabular}
\end{table*}
\subsection{Mobile app for data collection}

We developed an ESM tool named Waiting Time Activity tracker (hereinafter referred to as WTA). Participants installed this app on their own mobile phones through Google Play. The data was stored locally on their mobile phones; participants shared their data manually with the researchers (see below). Screenshots of WTA are shown in Fig. \ref{fig:wta}. The app is also available on GitHub\footnote{\url{https://github.com/jzheng23/WTA.git}}.

\begin{figure}
  \centering
  \includegraphics[width=\linewidth]{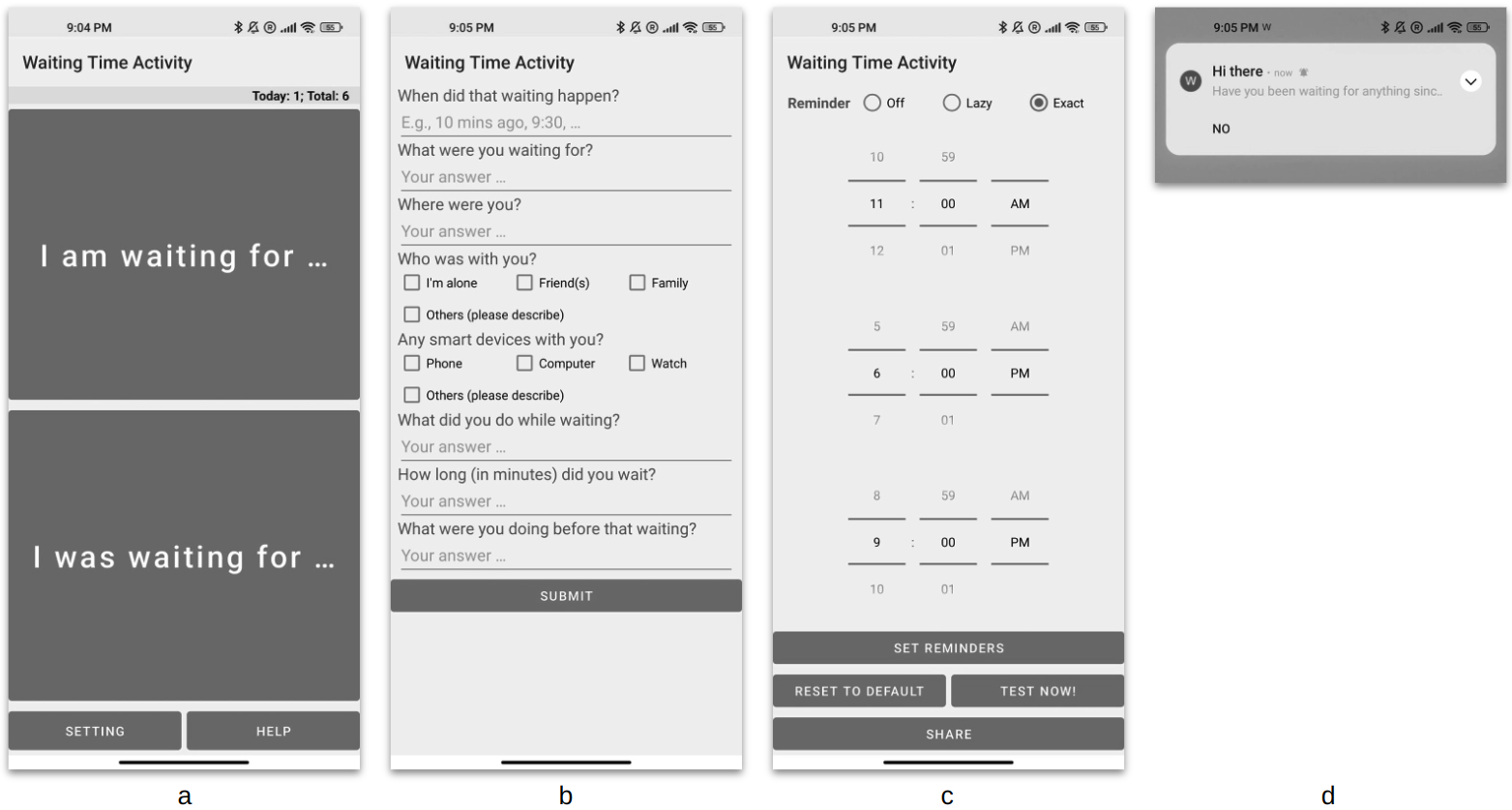}
  \caption{Screenshots of the WTA. (a) The main interface: participants can report an ongoing waiting by selecting "I am waiting for..." or report a previous waiting by selecting "I was waiting for ..." At the top is a tracker showing the number of reports made on the current day and the total number throughout the study; (b) The survey page: participants report their waiting time activities; (c) The settings page: participants can modify notifications and share their data file; (d) A notification: participants can tap this to make a report without manually open the app, ignore it, or dismiss it by tapping "NO". The screenshots were taken on a mobile phone using the dark system theme and recolored into negative black and white for readability.}
  \Description{This figure contains four images. Image a shows two large buttons "I am waiting for ..." and "I was waiting for ...", with two small button at the bottom "SETTING" and "HELP". At the top, a line of small text saying "Today: 1; Total: 6". Image b is the survey page with the following questions (and choices): When did that waiting happen? What were you waiting for? Where were you? Who was with you? Four checkbox: I'm alone, friends, family, others (please describe). Any smart devices with you? Four checkbox: Phone, computer, watch, others (please describe). What did you do while waiting? How long (in minutes) did you wait? What were you doing before that waiting? At the end is the "SUBMIT" button. Image c looks like an alarm setup page, with three time pickers showing the default settings: 11 a.m., 6 p.m., and 9 p.m. At the bottom were four small buttons: "SET REMINDERS", "RESET TO DEFAULT", "TEST NOW!", and "SHARE". Image d shows a notification with the title "Hi there" and body "Have you been waiting for anything since...". Below the message is a "NO" button.}
  \label{fig:wta}
\end{figure}

WTA collected participants’ responses with a combination of fixed-interval notifications and user-initiated reports. WTA sent three notifications each day. By default, notifications showed at 11 a.m., 6 p.m., and 9 p.m. and asked "Have you been waiting for anything since the last report?". Participants could set the notification time to their preference as long as there was no less than three hours between two notifications. The notification would stay active until either tapped, dismissed (by tapping the "No" button under the question), or replaced by the next notification. Participants could also disable the notification function.

Participants could make unlimited active, user-initiated reports at any time. Depending on whether they wanted to report something happening at that moment or that had already happened, they could select either "I am waiting for" or "I was waiting for", respectively. If they selected "I am waiting for", we would assume that they were waiting at that moment of reporting. If they selected "I was waiting for", we would ask explicitly when that waiting happened. Participants then reported their activity, devices available to them while waiting, and the location and duration of waiting.
  
At the end of each week of data collection, participants were asked to send their TXT data file by email with the first author. The procedure of data collection in WTA was transparent to participants. They could view and edit the file before sharing it.

\subsection{Procedure}

The ESM data collection continued for two weeks, which is the median study duration of previous ESM studies \cite{van-Berkel2018-wr}. After installing the application during an online 1-on-1 tutorial session with the first author, participants reported their waiting time activities for one week from that day—reporting did not need to start on a pre-specified day of the week. Participants were asked to make at least three reports each day to get fully paid. If a person did not meet the minimum requirement of 10 report submissions in the first week, they would not move forward to the next week of data collection but were paid according to the number of reports they had made. In addition, their data were deleted and removed from analysis. Participants who met the minimum requirement attended an online interview by the end of the first week, the purpose of which was to check if the software was working correctly. Eligible participants then continued to report their waiting time activities for another week. At the end of the second week, participants attended another online interview to share reflections on their own waiting time activities. Data were collected between June 2022 and June 2023.

We did not define "waiting" for the participants given that a definition at the conceptual level may appear abstract to participants and fail to assist their identification of waiting time in real life. Instead, we asked participants to identify waiting time against a list of representative scenarios stated in previous research \cite{Cai2015-sy, Cai2017-gc}, including waiting for an elevator to arrive, food to be delivered, water to boil, coffee to brew, the bus or taxi to arrive at its destination, a video to buffer, or a TV advertisement to end. The first author also explicitly mentioned that waiting not only involves waiting for something to happen; it also involves the case of waiting for someone. For example, a person may wait for their children to get ready to go to school every morning or their spouse to get ready to go to the gym.

\subsection{Data Analysis}

We coded self-reported activities into leisure, productive, and maintenance activities as in previous studies  \cite{Csikszentmihalyi1984-kj,Larson2001-pa,Skimina2020-kx}, with a reference to the ATUS Activity Coding Lexicons \cite{ATUS2023-jr}. Some participants reported more than one activity per report, but we only coded the first activity. Self-reported locations, following the coding practice suggested by prior research \cite{Hektner2006-ay}, were coded into home (including home office), work (e.g., office, schools, and universities), and other public places (e.g., restaurants and gyms). 

 We used R 4.3.1 for the data analysis, adopting the brms v2.20.1 package for regression analysis and the ggplot2 v3.4.2 for visualization. We built a codebook deductively with emphasis on the situational factors. We transcribed and analyzed the interview data following the procedure of thematic analysis \cite{Braun2006-yg}, aiming to gain insights that contextualize relevant quantitative findings. 

\section{Results}

Participants made 872 reports in total. Because people could make active reports even without being prompted, the response rate could exceed 100\%. The actual response rate ranged from 60\% (25 reports) to 179\% (75 reports); the median of 100\% is high compared with the average rate of 70\% in previous ESM studies \cite{van-Berkel2018-wr}. Out of all the reports, there were 720 active ones initiated by the participants themselves and 152 passive ones triggered by the preset prompts. Participants made 493 reports about ongoing waiting and 379 about previous waiting. For the latter, the median interval between waiting and making a report about it was 61 minutes. The reported duration of waiting was 18 minutes on average, ranging from 1 minute (e.g., waiting for water to boil) to 195 minutes (e.g., waiting for a baseball game to start), with a median of 11.5 minutes. The average total time spent waiting in any given day was 50 minutes; the median was 40 minutes. Appendix \ref{appendix:dur} details the distribution of participants’ waiting time across sessions and days. 

\subsection{RQ1: What Do People Do While Waiting?}
To answer RQ1, we calculated the total duration of each category and subcategory of activities. On average, participants reported spending 57.4\% of their waiting time on leisure activities, 22.5\% on productive activities, 17.1\% on maintenance activities, and 3.0\% without a clear purpose. Fig. \ref{fig:treemap} illustrates this part of the findings, supplemented by further details given in Appendix \ref{appendix:code}.

\begin{figure*}
  \centering
  \includegraphics[width=\linewidth]{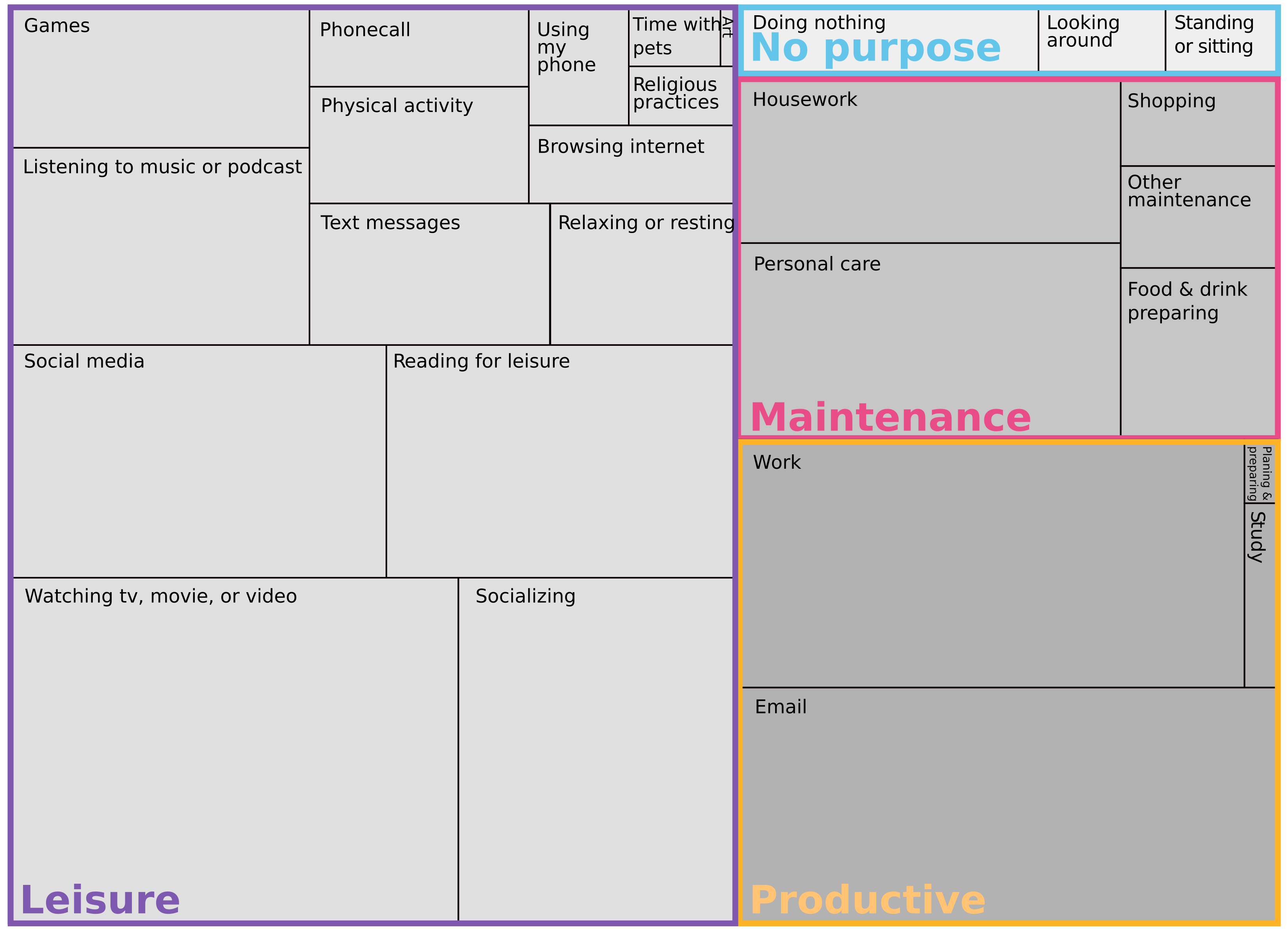}
  \caption{Activities while waiting. Areas indicate time spent in that sub-category of activities.}
  
  \Description{A treemap plot showing the allocation of waiting time among four categories: about 60\% on leisure, 20\% on productive, 20\% on maintenance activities, and a tiny portion on "No purpose". Within each category, there are sub-categories taking up areas according how much waiting time was spent on that sub-category. With a descending order, leisure activities include watching TV, movie, or video, socializing, social media, reading for leisure, listening to music or podcast, games, text messages, physical activity, relaxing or resting, phone-call, browsing internet, using my phone, religious practices, and time with pets. Productive activities include email, work, and study. Maintenance activities include personal care, housework, food and drink preparing, shopping, and other maintenance. Activities with no purpose include doing nothing, looking around, and standing or sitting.}
  \label{fig:treemap}
\end{figure*}

 Participants waiting time leisure activities included watching TV or videos, reading for leisure, using social media, listening to music, and socializing. Productive activities included work and study, checking and writing emails, and making plans. Since people can seldom complete a whole work project while waiting, people are more likely to continue or finish up previously started tasks, for example, "continue working on my project", or prepare for some tasks in the future, such as "getting ready for the meeting". Checking emails was the most common productive activities while waiting. Participants also reported maintenance activities while waiting, including doing household chores, taking care of oneself, and preparing to cook. Although sleeping and having meals are the most ubiquitous maintenance activities, participants rarely reported doing them while waiting. Drinking water, taking snacks, using the bathroom, and doing meditation were most often reported. Lastly, participants occasionally spent their waiting time without a clear purpose, such as just standing or sitting, looking around or staring at something, or, simply, doing “nothing.” 

 Overall, participants’ waiting time activities participants feature a high extent of fluidity. These activities must "fill opportunistic gaps, shrink, and expand until interrupted” \cite[p. 232]{Rattenbury2008-gz}. They are, by nature, distinctive from activities considered suitable for self-controlled and, sometimes, pre-planned work breaks \cite{Skatova2016-lx}. They also differ from activities that are common during the estimable time public transit takes to arrive at one’s destination \cite{Lee2023-xo}. 

\subsection{RQ2: How Do Situational Factors Affect People's Waiting Time Activities?}
Since participants spent about 60\% of their waiting time on leisure activities, we treated them as the reference to explore conditions under which people are more likely to do productive and maintenance activities. We built a multinomial logistic regression model for this investigation. The dependent variable was the category of waiting time activities (i.e., leisure, productive, or maintenance), using leisure as the baseline. The waiting time spent with “no purpose” (3\%) was excluded from this analysis. Independent variables in our final model included the duration of the waiting, whether the participants had access to their mobile phones, whether they had access to computers, whether they were at home, whether they were at workplaces, and whether the waiting occurred during lunch break (defined as 11:00 to 12:59 as in  \cite{Chen2023-dc}). We used lunch break instead of mealtime, as lunch break may arise from the blurring of work and non-work life during participants’ workdays\footnote{We also explored the effect of weekends and traditional office working hours (9 a.m. to 5 p.m.), but neither was significant. We did not include them in the final model because the model would, as a result of the small sample size, fail to converge given all the independent variables.}. Table \ref{table:regression} displays the results of the regression analysis. 

\begin{table*}
  \caption{Effects of Situational Factors on Waiting Time Activities}
  \label{table:regression}
\begin{tabular}{crrrrrrrr}
\toprule
& \multicolumn{4}{c}{Productive vs. Leisure} & \multicolumn{4}{c}{Maintenance vs. Leisure} \\
    \cmidrule(lr){2-5} \cmidrule(lr){6-9}
    {Variable} & {$B$} & {$SE\ B$} & {$t$} & {$p$} & {$B$} & {$SE\ B$} & {$t$} & {$p$} \\
    \midrule
(Intercept)               & -0.74                        & 0.29                          & -2.56                          & .01                            & -0.83                        & 0.28                          & -3.02                          & .001                           \\
Computer                  & 1.55                & 0.24                 & \textbf{6.51}                  & \textbf{\textless{}.001}       & -0.09                        & 0.24                          & -0.37                          & .35                            \\
Phone                     & 0.16                         & 0.51                          & 0.32                           & .37                            & -0.69               & 0.38                 & \textbf{-1.80}                 & \textbf{.04}                   \\
Workplace                 & 0.79                & 0.33                 & \textbf{2.42}                  & \textbf{.01}                   & 0.35                         & 0.45                          & 0.76                           & .22                            \\
Home                      & -0.23                        & 0.27                          & -0.85                          & .20                            & 1.08                & 0.25                 & \textbf{4.33}                  & \textbf{\textless{}.001}       \\
Lunchtime                 & 0.71                & 0.26                 & \textbf{2.71}                  & \textbf{.003}                  & 0.86                & 0.28                 & \textbf{3.04}                  & \textbf{.001}                  \\
Duration                  & 0.00                         & 0.00                          & 1.05                           & .15                            & 0.01                         & 0.01                          & 1.18                           & .12                            \\
 \bottomrule
    \addlinespace
    \multicolumn{9}{l}{\footnotesize *Bold font indicates statistical significance.}
\end{tabular}
\end{table*}

\textbf{\textit{Access to digital devices.}}
 Participants performed different waiting time activities depending on whether they had access to computers: access raised the probability of doing productive activities. When participants waited without computer access—accounting for approximately 59.5\% of waiting time—they spent 9.3\% of the time on productive activities, often checking email on their phones. 
 
 While waiting with computers, they spent 38.6\% of the time on productive activities, which themselves were more diverse, including "responding to emails", "working on my resume", or "working on my projects". Participants expressed awareness that they could do a greater variety of productive activities on their computers, whereas "there’s only so much [work] you could do on your cell phone” (P05). Some stated that, when the waiting time was long, they intentionally moved to their computers for work tasks. 

Different from computers,  mobile phones was almost universally accessible to our participants: they reported having a mobile phone 94.6\% of the time. When this was the case, participants spent 16.7\% of their waiting time on maintenance activities; without a phone, they spent 22.1\% of the time on these activities. Although participants could do leisure activities (e.g., watching TV), they were more likely to do maintenance activities (e.g., meditation) when without mobile phones. As P17 mentioned, "If I have my phone, I’m pretty much on social media. If I don’t have my phone, that’s my time to decompress distress through self-maintenance."

\begin{figure}
%  \centering
  \includegraphics[width=0.5\textwidth]{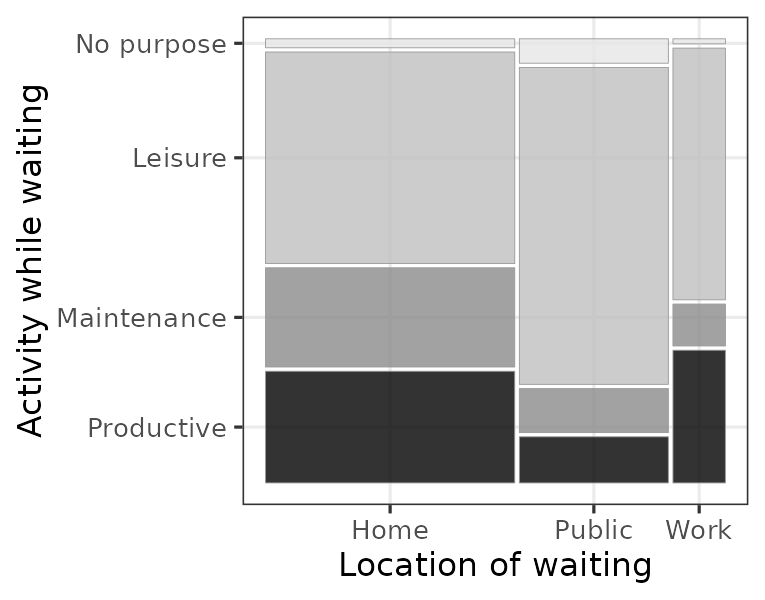}
  \caption{A mosaic plot of waiting time activities at different locations. Areas (and bar widths) indicate time spent in that activity category. More than half of the waiting time happened at home and maintenance activities were more likely to happen here. Only a small percentage of waiting time happened at workplaces, but productive activities were the most likely here. The proportion of leisure activities was the largest while waiting at public places.}
  \Description{A mosaic plot showing how waiting time is allocated among the four categories, which are leisure, maintenance, productive, and no purpose, at different location of waiting: home, work places, and public places. The ratio of home:public:work is roughly 6:3:1. Leisure takes up the most area in each location. Comparing across the three locations, leisure is largest at public places, maintenance is largest at home, and productive is largest at work places.}
  \label{fig:loc}
\end{figure}

\textbf{\textit{Location of waiting.}}
Participants’ waiting time activities varies according to the location where waiting occurred (Fig. \ref{fig:loc}). Half of participants’ waiting time occurred at home, one-third in public places, and about 10\% at workplaces. The high proportion of reported waiting at home could be partly explained by the fact that 12 participants work either remotely or hybridly. 

Participants performed a higher percentage of productive activities while waiting at workplaces (30.7\%) compared with home (25.6\%) and public places (10.6\%). They performed a higher percentage of maintenance activities while waiting at home (22.7\%) than at workplaces (9.7\%) or public places (10.1\%). Notably, it is possible that the effect of location on waiting time activities could be moderated by the participant’s working mode (i.e., working from home or not, see Appendix \ref{appendix:loc}). We did not test this potential moderating effect because the sample is not large enough to draw robust conclusions if split into further sub-groups.

\textbf{\textit{Lunchbreak routine.}}
Participants were less likely to do leisure activities (50.7\%) while waiting during lunch break compared with other waiting times (58.5\%). We speculate the effect of lunch break may arise from people’s preference of balancing activities of different natures. For example, P12 explained that he usually continued “working on some stuff while waiting for the microwave to heat the food up.” Previous research has documented similar cases where people leverage work-unrelated recurring slots (e.g., mealtime) to catch up on tasks left unfinished from work hours \cite{Skatova2016-lx}.

\textbf{\textit{Duration of waiting.}}
Our data reveal no clear relationship between the duration of waiting and types of activities performed\footnote{The duration data is not normally distributed. We have applied log transformation to the duration data and found that all the effects claimed in Table 2 would hold. Appendix \ref{appendix:regression} presents the results of logistic regression with log-transformed duration.}. This, again, supports the notion that waiting time is difficult to estimate and planned for proactively. As some of our participants mentioned, "Sometimes you don’t expect to wait a long time, and it ends up being a long time. So, you get caught off” (P21). As discussed in the Vierordt's law, that is, people often underestimate long time intervals and overestimate short ones, adding additional difficulty to time management \cite{Glasauer2021-xo,Lejeune2009-eu}. 

\section{Discussion}
The current study investigates how people spend their waiting time in daily lives. Our data showed that, on average, people allocated about 60\% of their waiting time on leisure activities, 20\% on productive activities, and 20\% on maintenance activities. People tend to perform different types of activities depending on the situation. They are more likely to do productive activities while waiting at workplaces and with computers and more likely to do maintenance activities while waiting at home, especially when without access to a mobile phone. They are less likely to do leisure activities when waiting occurs during a lunch break. Below, we discuss how these findings complement those reported in previous research as well as how they can inform future HCI work on the understanding and management of time use. 

\subsection{Leisure Activities as the Most Performed Waiting Time Activities}
The majority of waiting time in this study was spent on leisure activities (57.4\%), in contrast to the ATUS \cite{ATUS2022-qc}, which found that adults in the United States divide their overall daytime almost equally into leisure, productive, and maintenance activities. The dominance of leisure activities in our data challenges previous gap time research assumptions that people prefer to prioritize productivity \cite[e.g.,][]{Cai2017-gc, August2020-ew}. So, how should we interpret these contrasts? 

One possible interpretation is that people hope to spend their waiting time on productive and maintenance activities, but lack the opportunity to do so. Opportunity, in this context, refers to "all the factors that lie outside the individual that make the behavior possible or prompt it” \cite[p. 4]{Michie2011-nl}.  Our findings indicate that access to technology provides the opportunity for productive activities.

In particular, people tend to perform more productive activities when a computer is at hand. Having access to mobile phones, however, did not trigger more productive activities, perhaps due to a lack of professional software and a large screen for complex work \cite[e.g., ][]{Huang2009-sk, Shudong2005-qr}. Some of our data supports this. For instance, a few participants noted that, "it would be very hard to do it [writing a report] on the phone” (P06). However, checking emails, a task that can be easily managed on small devices, was the most frequently reported productive activity when participants had mobile phones during waiting time.

Location constitute the essential opportunity for maintenance activities. The predominant maintenance activities in our data (i.e., housework and food preparation) almost always take place at a person’s own home. Furthermore, many of these activities are performed a set number of times per day. For example, adults often have three meals and use the bathroom a finite number of times per day. In other words, there is a daily upper limit for many maintenance activities and, once that limit is reached, those activities will not be performed again in a waiting time later that same day.

Different from productive and maintenance activities, leisure activities can be performed with little situational constraint. Previous research suggested that people can habitually immerse themselves in social media consumption, as well as as well as other leisure activities, whenever they have access to a mobile phone \cite{Oulasvirta2012-en}. Approximately 95\% of the waiting time reported in our study satisfied this condition. Further, some of our participants reported that they switched back to leisure activities when the environment did not allow them to pursue productive or maintenance activities as they initially hoped.

It is also possible that people purposefully spend their waiting time on leisure activities despite the opportunity to perform productive or maintenance tasks. Several participants conveyed this sentiment in post-task reflections. For example, P02 shared that having "already worked the full day", he chose to "chill out and relax and just wait in the room, turn the CD on, sit there like a log." P21 said that on certain days he preferred to "be a little more relaxed and more inclined not to think about work as frequently." In short, people may purpose waiting time, especially during evenings and weekends, for a better maintenance of their boundaries between work and personal lives.

\subsection{Maintenance Activities as One Substantial Component of Waiting Time Activities}
Our data reveal a considerable proportion of waiting time spent doing maintenance activities, a finding seldom reported in previous gap time studies. We suspect that this discrepancy arises from varying operationalization of gap time across different studies. Our work complements previous literature by highlighting the ubiquitous aspect of gap time in everybody’s daily life.   
      
The bulk of previous research on gap time investigated how people spend work breaks in an office environment. For example, Mark and colleagues conducted a series of studies that indicate the positive connection between work breaks and stress management in high-tech workplaces \cite[e.g.,][]{Mark2014-yh, Mark2015-ad}. Skatova et al. found that most work breaks in offices were short \cite{Skatova2016-lx}. Kim et al. reported that office workers spent breaks more frequently on online activities rather than offline ones \cite{Kim2014-fk}. Across studies, participants mostly distributed their gap time across two categories of activities: either they leveraged the time to chase productivity (e.g., replying to an unanswered email) or they turned to leisure activities (e.g., watching videos via a digital device). The multiple types of maintenance activities captured in our data, such as personal care and meditation, may not fit the behavior norms of traditional workplaces, especially shared offices.

A small set of studies have operationalized gap time as transit time, investigating the time usage of commuters or train-riders \cite[e.g.,][]{Lee2023-xo, Lyons2005-rh, Keseru2018-bj, Axtell2008-kt}. As Lee et al. have pointed out, during transit time, people must cope with the instability of their physical and social surroundings \cite{Lee2023-xo}. Their choices of activities were also limited by the lack of privacy \cite{Axtell2008-kt}. It is, therefore, not surprising that few maintenance activities have been observed and reported in this research context.

Our current research centers around waiting time, a more general format of gap time that occurs ubiquitously. By detaching gap time from preset scenarios (e.g., office work, transportation), we gain the opportunity to capture various types of maintenance activities that people perform in between other scheduled events. A growing number of recent HCI studies have demonstrated that a relentless pursuit of productivity is potentially counterproductive in the long term \cite[e.g.,][]{Isham2021-tw}. People are increasingly encouraged to reflect on their time usage with a view to optimizing their well-being \cite{Guillou2020-sb, Odom2018-dv, Odom2022-aj}. Our work contributes to the broad spectrum of literature that considers well-being as being equally important to, if not more important than, productivity \cite{Guillou2020-sb, Leshed2011-zp, Ruth_Eikhof2007-pl, Schor2008-mq, Ozmen_Garibay2023-ct, Stephanidis2019-hl}. Allocating sufficient time to non-productive activities enables people to sustain their energy \cite{Albulescu2022-nn, Clark2000-az}.

Moreover, it is reasonable to argue that, for people who choose to leverage their waiting time for non-productive purposes, maintenance activities may elicit less negative self-criticism than leisure activities. The latter often induce feelings of absent-mindedness, meaningless, and regret, rather than sustaining a person’s positive self-image \cite{Terzimehic2022-si, Lukoff2018-ws, Cho2021-xe}. Excessive use of phones for leisure can be harmful for mental health and professional performance \cite{Busch2021-fi, Ochs2022-fw, Tandon2022-ki}. In contrast, engaging in maintenance activities enables people to temporarily distance themselves from work, while avoiding digital game addiction or the rabbit hole of internet browsing \cite{Terzimehic2023-yi}.  

\subsection{Implications for the Management of Waiting Time}
Findings of our study suggest waiting time is an essential modality for balancing different activities and shed light on the design of behavioral protocols and technical tools for the management of waiting time. Below, we outline three implications of our research.

\textbf{\textit{Waiting time usage that reflects different values.}} As our data suggests, people naturally spend their waiting time on various categories of activities. Some of those activities promote productivity, while others contribute to other valuable aspects of a person’s well-being. One straightforward implication for future technology design is to assist people to stay aware of this diverse set of possibilities, as well as the underlying values attaching to each possibility. Such awareness would prevent individuals from making productivity the solo focus of their attention \cite[e.g.,][]{Gouko2017-nh,Lane2007-jn,Zeidan2010-ps}.

We envision a design space where future self-tracking tools, such as an upgraded version of our WTA app, could recommend a variety of activities to people waiting. If an app user has previous recorded work-relevant waiting time activities, a higher proportion of maintenance and leisure activities could be displayed in upcoming recommendations. Activity reports documented in our research creates a repository from which these recommendations can be derived. This diversity of options enables users to switch between different tasks, contributing to sustainable life and work \cite{Steup2022-gx}. 

\textbf{\textit{Waiting time usage that leverages the situational opportunities.}} People are likely to value certain goals over others when managing a specific instance of waiting. Thus, future technologies should consider the management of waiting time activities according to situational factors. For instance, our current research found a strong association between waiting time activities and devices available: waiting time activities are more likely to be productive when there is access to a computer, but people turn to leisure activities when waiting with mobile phones. Skatova et al. \cite{Skatova2016-lx} also suggested that digital ecosystems should make use of multiple devices to balance work and break activities. Therefore, an app could prompt a switch from a mobile device to a computer for those who intend to leverage their waiting time to get work done. If an individual prefers to relax during waiting time, however, an app could advise them to move away from their computer to decrease exposure to incoming tasks.

Notably, previous HCI research has assumed a natural fit among waiting time, mobile devices, and microlearning or microtasking for productivity \cite[e.g.,][]{Cai2017-gc, Iqbal2018-tj, Williams2019-ls}. We caution future system designers to reconsider this oversimplified assumption. Our data indicate that mobile phones might not be conductive to tasks requiring complex input and a high level of concentration from the waiting person. Therefore, microlearning or microtasking tools should allow people to engage in tasks of various formats, such as recognition-centered tasks on a mobile phone and typing-and-recalling tasks on a computer \cite{Haist1992-nv}.

\textbf{\textit{Waiting time usage through pre-planning.}} Another important finding of our research is that people are not adept at predicting their waiting time, nor can they always devise prior activity plans for that time. Our data indicate no statistical relationship between the duration of waiting and the activities engaged in by our participants. Qualitative feedback from participants also confirmed that they sometimes made incorrect estimations about their waiting time and, consequently, delayed the adjustment of activities to perform. Future time management tools should help people achieve a greater sense of control over this plastic part of their daily schedule.

While the current dataset does not permit us to draw detailed conclusions, it signals the potential to capture characteristic waiting scenarios at a given location or recurring patterns of waiting across different occasions for the same individual. An expanded deployment of EMS tools, such as the WTA app, would permit HCI researchers to gain more comprehensive data and develop predictive models aiding in the pre-planning of waiting time usage.

\subsection{Limitations and Directions for Future Studies}
Our research is subject to certain limitations related to our sampling strategy and data collection method. In particular, we exclusively recruited adult Android users, working in the United States. Future studies should consider recruiting both Android and iOS users, as well as participants with a more diverse set of backgrounds. Our WTA app only supports manual text input for open-ended questions, although voice-to-text is possible through tools available on the Android platform. Future studies may consider enabling voice input to produce audio files and allowing images or videos to be uploaded to increase the richness of data \cite[e.g.,][]{Luo2022-qm}. Also, for waiting time that had already been completed, participants only reported the activities they actually performed, putting aside the question of how they intended to spend that time. Previous studies have suggested that people are likely to engage in harmful activities, such as absent-mined phone usage \cite{Haliburton2022-ch,Terzimehic2022-si}, when encountering unexpected gap time. Future research should explicitly consider people’s intentions for waiting time activities, as well as situational factors that facilitate or hinder the translation of these intentions into the actual activities being performed. This approach ensures that our technology design can align with user preferences and enhance their waiting time experience.

Moreover, while our research uncovered the influence of situational factors, such as location and device availability, on people’s waiting time activities, these factors are not fully independent. For example, people usually have access to their computers at the workplace, but not in other public places. The association between at-home waiting and maintenance activities appears stable, even after splitting participants into different subgroups (e.g., remote workers vs. others); however, the relationship between at-home waiting and productive activities appears more complex across subgroups of participants. Despite our interest in examining these nuances, the current dataset does not contain sufficient information for a thorough exploration. Future research involving more participants or with a focus on some specific interactions are necessary to obtain relevant insights.

Last but not least, we acknowledge that the use of ESM as our primary method has introduced potential limitations to the current research. For instance, it is hard for participants to file reports detailing a sequence of activities performed over the same waiting period. Also, the time spent with no purpose might be underestimated, because participants’ attention is directed to what they did, not to the absence of specific activities. We carefully reflected on our methodological choices: our comparison of ESM with other HCI research methods aided us in making sense of the strengths and weakness of ESM in the current research context (see Table \ref{table:methods} for a summary).

\begin{table*}
    \caption{A Comparison Between ESM and Other Methods in Terms of Their Strengths and Weaknesses for Empirical Research}
    \label{table:methods}
    \begin{tabular}{lccccc}
    \toprule
    Attributes \textbackslash{ Method} & ESM & Shadowing & Wearable cameras & Survey & Passive tracking  \\
    \midrule
    Recording situational factors & Yes & Yes & No & No & Yes  \\ 
    Recognizing physical activities & Yes & Yes & Yes & Yes & Limited  \\ 
    Identifying waiting from first-person accounts & Yes & No & No & Yes & No  \\
    Being intrusive & No & Yes & Yes & No & No  \\ 
    Raising privacy concerns & No & Yes & Yes & No & Yes  \\ 
    Inducing self-report bias & Yes & No & No & Yes & No  \\ 
    Requiring effort from participants & Yes & No & No & Yes & No  \\ 
    Number of participants commonly involved & Small & Small & Small & Large & Medium \\
    \midrule
    Representative example of this kind & This study & \cite{Chen2023-dc} & \cite{Skatova2016-lx} & \cite{ATUS2022-qc} & \cite{Kang2017-cf}  \\ 
    \bottomrule
    \end{tabular}
\end{table*}

Specifically, previous time gap studies collected data with shadowing observation \cite{Lee2023-xo} or wearable cameras (e.g., Autographer \cite{Skatova2016-lx}). Although these methods do not rely on participants’ self-reports as ESM does, they are more intrusive. Participants may alter their behavior knowing they are being observed or recorded or due to privacy concerns. The number of participants involved in this area of research is usually small \cite[e.g.,][]{Skatova2016-lx,Lee2023-xo}.

Time use surveys, on the other hand, can obtain data from a much larger number of participants. For example, ATUS \cite{ATUS2022-qc} collects responses from thousands of individuals per year. Their datasets support more complex models and are ideal for exploring individual differences. However, since they rely on self-reported data based on participants’ long-term memory recall, their accuracy may not be as high as ESM reports collected in-situ. Besides, time use surveys do not record situational information such as the time and location of each reported activity.

More recent ESM studies often incorporate mobile phone and computer usage logging data as additional data streams \cite[e.g.,][]{Lukoff2023-lt,Zhang2022-rm}. The logging data can record the sequence of activities accurately in milliseconds with little burden on the participants. Nevertheless, physical activities occurring outside of mobile phones or computers cannot be captured through logging data and privacy concerns may arise. If not combined with ESM reports, relying on passive data collection alone faces the challenge of detecting waiting time through physical sensing. The Zaturi project \cite{Kang2017-cf}, for example, adopted mobile phone sensors for the auto-detection of tiny fragments of free time. The detection is set to occur immediately after a person has finished or cancelled a mobile phone call, when the person has performed extended use (90 seconds) of social media, or while the person is walking and listening to music through headphones. Despite the convenience of implementing such a sensing system against requesting people’s self-reports, the authors noted that it is extremely challenging and impractical to detect all types of micro spare time and at every occurrence. The above reflections remind us to plan future research with a combined but meticulously selected set of methods.

\section{Conclusion}
People allocate roughly 60\% of their waiting time to leisure activities, 20\% to productive activities, and 20\% to maintenance activities. These patterns are significantly influenced by situational factors. People are more likely to do productive activities while waiting at workplaces and with access to computers. They are more likely to do maintenance activities while waiting at home and without access to mobile phones. They are also less likely to do leisure activities while waiting during a lunch break. Instead of pursuing productivity relentlessly, we should empower users to choose how they spend their waiting time, taking into account their specific situations and preferences. This user-centric approach ensures a more balanced and flexible utilization of waiting time.

\begin{acks}
We thank the anonymous reviewers for their valuable comments on this paper. We also thank Janel Rana for her help with recruitment and Eun Kyoung Choe, Leo Zhicheng Liu, Ruipu Hu, Yimin Xiao, and Yongle Zhang for their feedback at the early stages of this research. 
\end{acks}
\bibliographystyle{ACM-Reference-Format}
\bibliography{references}


\begin{thebibliography}{68}

%%% the Web address.

\ifx \showCODEN    \undefined \def \showCODEN     #1{\unskip}     \fi
\ifx \showDOI      \undefined \def \showDOI       #1{#1}\fi
\ifx \showISBNx    \undefined \def \showISBNx     #1{\unskip}     \fi
\ifx \showISBNxiii \undefined \def \showISBNxiii  #1{\unskip}     \fi
\ifx \showISSN     \undefined \def \showISSN      #1{\unskip}     \fi
\ifx \showLCCN     \undefined \def \showLCCN      #1{\unskip}     \fi
\ifx \shownote     \undefined \def \shownote      #1{#1}          \fi
\ifx \showarticletitle \undefined \def \showarticletitle #1{#1}   \fi
\ifx \showURL      \undefined \def \showURL       {\relax}        \fi
% invisible to TeX
\providecommand\bibfield[2]{#2}
\providecommand\bibinfo[2]{#2}
\providecommand\natexlab[1]{#1}
\providecommand\showeprint[2][]{arXiv:#2}

\bibitem[Albulescu et~al\mbox{.}(2022)]%
        {Albulescu2022-nn}
\bibfield{author}{\bibinfo{person}{Patricia Albulescu}, \bibinfo{person}{Irina Macsinga}, \bibinfo{person}{Andrei Rusu}, \bibinfo{person}{Coralia Sulea}, \bibinfo{person}{Alexandra Bodnaru}, {and} \bibinfo{person}{Bogdan~Tudor Tulbure}.} \bibinfo{year}{2022}\natexlab{}.
\newblock \showarticletitle{``Give me a break!'' A systematic review and meta-analysis on the efficacy of micro-breaks for increasing well-being and performance}.
\newblock \bibinfo{journal}{\emph{PLoS One}} \bibinfo{volume}{17}, \bibinfo{number}{8} (\bibinfo{date}{Aug.} \bibinfo{year}{2022}), \bibinfo{pages}{e0272460}.
\newblock
\showISSN{1932-6203}
\urldef\tempurl%
\url{https://doi.org/10.1371/journal.pone.0272460}
\showDOI{\tempurl}


\bibitem[August et~al\mbox{.}(2020)]%
        {August2020-ew}
\bibfield{author}{\bibinfo{person}{Tal August}, \bibinfo{person}{Shamsi~T Iqbal}, \bibinfo{person}{Michael Gamon}, {and} \bibinfo{person}{Mark Encarnaci{\'o}n}.} \bibinfo{year}{2020}\natexlab{}.
\newblock \showarticletitle{Characterizing the Mobile Microtask Writing Process}. In \bibinfo{booktitle}{\emph{22nd International Conference on {Human-Computer} Interaction with Mobile Devices and Services}} (Oldenburg, Germany) \emph{(\bibinfo{series}{MobileHCI '20})}. \bibinfo{publisher}{Association for Computing Machinery}, \bibinfo{address}{New York, NY, USA}, \bibinfo{pages}{1--12}.
\newblock
\showISBNx{9781450375160}
\urldef\tempurl%
\url{https://doi.org/10.1145/3379503.3403541}
\showDOI{\tempurl}


\bibitem[Axtell et~al\mbox{.}(2008)]%
        {Axtell2008-kt}
\bibfield{author}{\bibinfo{person}{Carolyn Axtell}, \bibinfo{person}{Donald Hislop}, {and} \bibinfo{person}{Steve Whittaker}.} \bibinfo{year}{2008}\natexlab{}.
\newblock \showarticletitle{Mobile technologies in mobile spaces: Findings from the context of train travel}.
\newblock \bibinfo{journal}{\emph{Int. J. Hum. Comput. Stud.}} \bibinfo{volume}{66}, \bibinfo{number}{12} (\bibinfo{date}{Dec.} \bibinfo{year}{2008}), \bibinfo{pages}{902--915}.
\newblock


\bibitem[Braun and Clarke(2006)]%
        {Braun2006-yg}
\bibfield{author}{\bibinfo{person}{Virginia Braun} {and} \bibinfo{person}{Victoria Clarke}.} \bibinfo{year}{2006}\natexlab{}.
\newblock \showarticletitle{Using thematic analysis in psychology}.
\newblock \bibinfo{journal}{\emph{Qual. Res. Psychol.}} \bibinfo{volume}{3}, \bibinfo{number}{2} (\bibinfo{date}{Jan.} \bibinfo{year}{2006}), \bibinfo{pages}{77--101}.
\newblock
\showISSN{1478-0887, 1478-0895}
\urldef\tempurl%
\url{https://doi.org/10.1191/1478088706qp063oa}
\showDOI{\tempurl}


\bibitem[Busch and McCarthy(2021)]%
        {Busch2021-fi}
\bibfield{author}{\bibinfo{person}{Peter~André Busch} {and} \bibinfo{person}{Stephen McCarthy}.} \bibinfo{year}{2021}\natexlab{}.
\newblock \showarticletitle{Antecedents and consequences of problematic smartphone use: A systematic literature review of an emerging research area}.
\newblock \bibinfo{journal}{\emph{Comput. Human Behav.}} \bibinfo{volume}{114}, \bibinfo{number}{106414} (\bibinfo{date}{Jan.} \bibinfo{year}{2021}), \bibinfo{pages}{106414}.
\newblock


\bibitem[Cai(2015)]%
        {Cai2015-sy}
\bibfield{author}{\bibinfo{person}{Carrie~J Cai}.} \bibinfo{year}{2015}\natexlab{}.
\newblock \showarticletitle{Wait-learning: Leveraging wait time for education}. In \bibinfo{booktitle}{\emph{Adjunct Proceedings of the 28th Annual {ACM} Symposium on User Interface Software \& Technology}} (Daegu Kyungpook Republic of Korea) \emph{(\bibinfo{series}{UIST '15 Adjunct})}. \bibinfo{publisher}{ACM}, \bibinfo{address}{New York, NY, USA}, \bibinfo{pages}{13--16}.
\newblock
\showISBNx{9781450337809}
\urldef\tempurl%
\url{https://doi.org/10.1145/2815585.2815589}
\showDOI{\tempurl}


\bibitem[Cai et~al\mbox{.}(2017)]%
        {Cai2017-gc}
\bibfield{author}{\bibinfo{person}{Carrie~J Cai}, \bibinfo{person}{Anji Ren}, {and} \bibinfo{person}{Robert~C Miller}.} \bibinfo{year}{2017}\natexlab{}.
\newblock \showarticletitle{{WaitSuite}: Productive use of diverse waiting moments}.
\newblock \bibinfo{journal}{\emph{ACM Trans. Comput. Hum. Interact.}} \bibinfo{volume}{24}, \bibinfo{number}{1} (\bibinfo{date}{Feb.} \bibinfo{year}{2017}), \bibinfo{pages}{1--41}.
\newblock
\showISSN{1073-0516, 1557-7325}
\urldef\tempurl%
\url{https://doi.org/10.1145/3044534}
\showDOI{\tempurl}


\bibitem[Cao et~al\mbox{.}(2021)]%
        {Cao2021-ao}
\bibfield{author}{\bibinfo{person}{Hancheng Cao}, \bibinfo{person}{Chia-Jung Lee}, \bibinfo{person}{Shamsi Iqbal}, \bibinfo{person}{Mary Czerwinski}, \bibinfo{person}{Priscilla N~Y Wong}, \bibinfo{person}{Sean Rintel}, \bibinfo{person}{Brent Hecht}, \bibinfo{person}{Jaime Teevan}, {and} \bibinfo{person}{Longqi Yang}.} \bibinfo{year}{2021}\natexlab{}.
\newblock \showarticletitle{Large Scale Analysis of Multitasking Behavior During Remote Meetings}. In \bibinfo{booktitle}{\emph{Proceedings of the 2021 {CHI} Conference on Human Factors in Computing Systems}} (Yokohama, Japan) \emph{(\bibinfo{series}{CHI '21}, \bibinfo{number}{Article 448})}. \bibinfo{publisher}{Association for Computing Machinery}, \bibinfo{address}{New York, NY, USA}, \bibinfo{pages}{1--13}.
\newblock
\showISBNx{9781450380966}
\urldef\tempurl%
\url{https://doi.org/10.1145/3411764.3445243}
\showDOI{\tempurl}


\bibitem[Chen et~al\mbox{.}(2023)]%
        {Chen2023-dc}
\bibfield{author}{\bibinfo{person}{Yu-Chun Chen}, \bibinfo{person}{Yu-Jen Lee}, \bibinfo{person}{Kuei-Chun Kao}, \bibinfo{person}{Jie Tsai}, \bibinfo{person}{En-Chi Liang}, \bibinfo{person}{Wei-Chen Chiu}, \bibinfo{person}{Faye Shih}, {and} \bibinfo{person}{Yung-Ju Chang}.} \bibinfo{year}{2023}\natexlab{}.
\newblock \showarticletitle{Are you killing time? Predicting smartphone users' time-killing moments via fusion of smartphone sensor data and screenshots}. In \bibinfo{booktitle}{\emph{Proceedings of the 2023 {CHI} Conference on Human Factors in Computing Systems}} (Hamburg Germany) \emph{(\bibinfo{series}{CHI '23}, \bibinfo{number}{Article 647})}. \bibinfo{publisher}{ACM}, \bibinfo{address}{New York, NY, USA}, \bibinfo{pages}{1--19}.
\newblock
\showISBNx{9781450394215}
\urldef\tempurl%
\url{https://doi.org/10.1145/3544548.3580689}
\showDOI{\tempurl}


\bibitem[Cho et~al\mbox{.}(2021)]%
        {Cho2021-xe}
\bibfield{author}{\bibinfo{person}{Hyunsung Cho}, \bibinfo{person}{Daeun Choi}, \bibinfo{person}{Donghwi Kim}, \bibinfo{person}{Wan~Ju Kang}, \bibinfo{person}{Eun~Kyoung Choe}, {and} \bibinfo{person}{Sung-Ju Lee}.} \bibinfo{year}{2021}\natexlab{}.
\newblock \showarticletitle{Reflect, not regret: Understanding regretful smartphone use with app feature-level analysis}.
\newblock \bibinfo{journal}{\emph{Proc. ACM Hum. Comput. Interact.}} \bibinfo{volume}{5}, \bibinfo{number}{CSCW2} (\bibinfo{date}{Oct.} \bibinfo{year}{2021}), \bibinfo{pages}{1--36}.
\newblock
\showISSN{2573-0142}
\urldef\tempurl%
\url{https://doi.org/10.1145/3479600}
\showDOI{\tempurl}


\bibitem[Clark(2000)]%
        {Clark2000-az}
\bibfield{author}{\bibinfo{person}{Sue~Campbell Clark}.} \bibinfo{year}{2000}\natexlab{}.
\newblock \showarticletitle{Work/family border theory: A new theory of work/family balance}.
\newblock \bibinfo{journal}{\emph{Hum. Relat.}} \bibinfo{volume}{53}, \bibinfo{number}{6} (\bibinfo{date}{June} \bibinfo{year}{2000}), \bibinfo{pages}{747--770}.
\newblock
\showISSN{0018-7267, 1741-282X}
\urldef\tempurl%
\url{https://doi.org/10.1177/0018726700536001}
\showDOI{\tempurl}


\bibitem[Csikszentmihalyi and Larson(1984)]%
        {Csikszentmihalyi1984-kj}
\bibfield{author}{\bibinfo{person}{Mihaly Csikszentmihalyi} {and} \bibinfo{person}{Reed Larson}.} \bibinfo{year}{1984}\natexlab{}.
\newblock \bibinfo{booktitle}{\emph{Being Adolescent: Conflict and Growth during the Teenage Years}}.
\newblock \bibinfo{publisher}{Basic Books}, \bibinfo{address}{London, England}.
\newblock
\showISSN{0094-3061, 1939-8638}


\bibitem[Gershuny et~al\mbox{.}(2023)]%
        {MTUS2023-os}
\bibfield{author}{\bibinfo{person}{Jonathan Gershuny}, \bibinfo{person}{Marga Vega-Rapun}, {and} \bibinfo{person}{Lamote Juana}.} \bibinfo{year}{2023}\natexlab{}.
\newblock \bibinfo{title}{Multinational Time Use Study}.
\newblock \bibinfo{howpublished}{\url{https://www.timeuse.org/mtus}}.
\newblock
\newblock
\shownote{Accessed: 2023-12-10}.


\bibitem[Glasauer and Shi(2021)]%
        {Glasauer2021-xo}
\bibfield{author}{\bibinfo{person}{Stefan Glasauer} {and} \bibinfo{person}{Zhuanghua Shi}.} \bibinfo{year}{2021}\natexlab{}.
\newblock \showarticletitle{The origin of Vierordt's law: The experimental protocol matters}.
\newblock \bibinfo{journal}{\emph{PsyCh J.}} \bibinfo{volume}{10}, \bibinfo{number}{5} (\bibinfo{date}{Oct.} \bibinfo{year}{2021}), \bibinfo{pages}{732--741}.
\newblock
\showISSN{2046-0260, 2046-0252}
\urldef\tempurl%
\url{https://doi.org/10.1002/pchj.464}
\showDOI{\tempurl}


\bibitem[Gouko and Arakawa(2017)]%
        {Gouko2017-nh}
\bibfield{author}{\bibinfo{person}{Manabu Gouko} {and} \bibinfo{person}{Yuka Arakawa}.} \bibinfo{year}{2017}\natexlab{}.
\newblock \showarticletitle{A Coaster Robot that Encourages Office Workers to Drink Water}. In \bibinfo{booktitle}{\emph{Proceedings of the 5th International Conference on Human Agent Interaction}} (Bielefeld, Germany) \emph{(\bibinfo{series}{HAI '17})}. \bibinfo{publisher}{Association for Computing Machinery}, \bibinfo{address}{New York, NY, USA}, \bibinfo{pages}{447--449}.
\newblock
\showISBNx{9781450351133}
\urldef\tempurl%
\url{https://doi.org/10.1145/3125739.3132584}
\showDOI{\tempurl}


\bibitem[Guillou et~al\mbox{.}(2020)]%
        {Guillou2020-sb}
\bibfield{author}{\bibinfo{person}{Hayley Guillou}, \bibinfo{person}{Kevin Chow}, \bibinfo{person}{Thomas Fritz}, {and} \bibinfo{person}{Joanna McGrenere}.} \bibinfo{year}{2020}\natexlab{}.
\newblock \showarticletitle{Is your time well spent? Reflecting on knowledge work more holistically}. In \bibinfo{booktitle}{\emph{Proceedings of the 2020 {CHI} Conference on Human Factors in Computing Systems}} (Honolulu HI USA) \emph{(\bibinfo{series}{CHI '20})}. \bibinfo{publisher}{ACM}, \bibinfo{address}{New York, NY, USA}, \bibinfo{pages}{1--9}.
\newblock
\showISBNx{9781450367080}
\urldef\tempurl%
\url{https://doi.org/10.1145/3313831.3376586}
\showDOI{\tempurl}


\bibitem[Hahn et~al\mbox{.}(2019)]%
        {Hahn2019-zy}
\bibfield{author}{\bibinfo{person}{Nathan Hahn}, \bibinfo{person}{Shamsi~T Iqbal}, {and} \bibinfo{person}{Jaime Teevan}.} \bibinfo{year}{2019}\natexlab{}.
\newblock \showarticletitle{Casual Microtasking: Embedding Microtasks in Facebook}. In \bibinfo{booktitle}{\emph{Proceedings of the 2019 {CHI} Conference on Human Factors in Computing Systems}} (Glasgow Scotland Uk) \emph{(\bibinfo{series}{CHI '19}, \bibinfo{number}{Paper 19})}. \bibinfo{publisher}{ACM}, \bibinfo{address}{New York, NY, USA}, \bibinfo{pages}{1--9}.
\newblock
\showISBNx{9781450359702}
\urldef\tempurl%
\url{https://doi.org/10.1145/3290605.3300249}
\showDOI{\tempurl}


\bibitem[Haist et~al\mbox{.}(1992)]%
        {Haist1992-nv}
\bibfield{author}{\bibinfo{person}{Frank Haist}, \bibinfo{person}{Arthur~P Shimamura}, {and} \bibinfo{person}{Larry~R Squire}.} \bibinfo{year}{1992}\natexlab{}.
\newblock \showarticletitle{On the relationship between recall and recognition memory}.
\newblock \bibinfo{journal}{\emph{J. Exp. Psychol. Learn. Mem. Cogn.}} \bibinfo{volume}{18}, \bibinfo{number}{4} (\bibinfo{year}{1992}), \bibinfo{pages}{691--702}.
\newblock
\showISSN{0278-7393, 1939-1285}
\urldef\tempurl%
\url{https://doi.org/10.1037/0278-7393.18.4.691}
\showDOI{\tempurl}


\bibitem[Haliburton et~al\mbox{.}(2022)]%
        {Haliburton2022-ch}
\bibfield{author}{\bibinfo{person}{Luke Haliburton}, \bibinfo{person}{Maximilian Lammel}, \bibinfo{person}{Jakob Karolus}, {and} \bibinfo{person}{Albrecht Schmidt}.} \bibinfo{year}{2022}\natexlab{}.
\newblock \showarticletitle{Think inside the box: Investigating the consequences of everyday physical opt-out strategies for mindful smartphone use}. In \bibinfo{booktitle}{\emph{Proceedings of the 21st International Conference on Mobile and Ubiquitous Multimedia}} (Lisbon Portugal) \emph{(\bibinfo{series}{MUM '22})}. \bibinfo{publisher}{ACM}, \bibinfo{address}{New York, NY, USA}, \bibinfo{pages}{37--46}.
\newblock
\showISBNx{9781450398206}
\urldef\tempurl%
\url{https://doi.org/10.1145/3568444.3568452}
\showDOI{\tempurl}


\bibitem[Hektner et~al\mbox{.}(2006)]%
        {Hektner2006-ay}
\bibfield{author}{\bibinfo{person}{Joel~M Hektner}, \bibinfo{person}{Jennifer~A Schmidt}, {and} \bibinfo{person}{Mihaly Csikszentmihalyi}.} \bibinfo{year}{2006}\natexlab{}.
\newblock \bibinfo{booktitle}{\emph{Experience sampling method: Measuring the quality of everyday life}}.
\newblock \bibinfo{publisher}{SAGE Publications}, \bibinfo{address}{Thousand Oaks, CA}.
\newblock
\showISBNx{9781412949231, 9781412984201}
\urldef\tempurl%
\url{https://doi.org/10.4135/9781412984201}
\showDOI{\tempurl}


\bibitem[Huang(2009)]%
        {Huang2009-sk}
\bibfield{author}{\bibinfo{person}{Kuo-Ying Huang}.} \bibinfo{year}{2009}\natexlab{}.
\newblock \showarticletitle{Challenges in human-computer interaction design for mobile devices}. In \bibinfo{booktitle}{\emph{Proceedings of the World Congress on Engineering and Computer Science}}, Vol.~\bibinfo{volume}{1}. \bibinfo{publisher}{Newswood Limited}, \bibinfo{address}{Hong Kong}, \bibinfo{pages}{1--6}.
\newblock
\urldef\tempurl%
\url{http://iaeng.org/publication/WCECS2009/WCECS2009_pp236-241.pdf}
\showURL{%
\tempurl}


\bibitem[Iqbal et~al\mbox{.}(2018)]%
        {Iqbal2018-tj}
\bibfield{author}{\bibinfo{person}{Shamsi~T Iqbal}, \bibinfo{person}{Jaime Teevan}, \bibinfo{person}{Dan Liebling}, {and} \bibinfo{person}{Anne~Loomis Thompson}.} \bibinfo{year}{2018}\natexlab{}.
\newblock \showarticletitle{Multitasking with play write, a mobile microproductivity writing tool}. In \bibinfo{booktitle}{\emph{Proceedings of the 31st Annual {ACM} Symposium on User Interface Software and Technology}} (Berlin Germany) \emph{(\bibinfo{series}{UIST '18})}. \bibinfo{publisher}{ACM}, \bibinfo{address}{New York, NY, USA}, \bibinfo{pages}{411--422}.
\newblock
\showISBNx{9781450359481}
\urldef\tempurl%
\url{https://doi.org/10.1145/3242587.3242611}
\showDOI{\tempurl}


\bibitem[Isham et~al\mbox{.}(2021)]%
        {Isham2021-tw}
\bibfield{author}{\bibinfo{person}{Amy Isham}, \bibinfo{person}{Simon Mair}, {and} \bibinfo{person}{Tim Jackson}.} \bibinfo{year}{2021}\natexlab{}.
\newblock \showarticletitle{Worker wellbeing and productivity in advanced economies: Re-examining the link}.
\newblock \bibinfo{journal}{\emph{Ecol. Econ.}} \bibinfo{volume}{184}, \bibinfo{number}{106989} (\bibinfo{date}{June} \bibinfo{year}{2021}), \bibinfo{pages}{106989}.
\newblock
\showISSN{0921-8009, 1873-6106}
\urldef\tempurl%
\url{https://doi.org/10.1016/j.ecolecon.2021.106989}
\showDOI{\tempurl}


\bibitem[Kang et~al\mbox{.}(2017)]%
        {Kang2017-cf}
\bibfield{author}{\bibinfo{person}{Bumsoo Kang}, \bibinfo{person}{Chulhong Min}, \bibinfo{person}{Wonjung Kim}, \bibinfo{person}{Inseok Hwang}, \bibinfo{person}{Chunjong Park}, \bibinfo{person}{Seungchul Lee}, \bibinfo{person}{Sung-Ju Lee}, {and} \bibinfo{person}{Junehwa Song}.} \bibinfo{year}{2017}\natexlab{}.
\newblock \showarticletitle{Zaturi: We put together the 25th hour for you. Create a book for your baby}. In \bibinfo{booktitle}{\emph{Proceedings of the 2017 {ACM} Conference on Computer Supported Cooperative Work and Social Computing}} (Portland Oregon USA) \emph{(\bibinfo{series}{CSCW '17})}. \bibinfo{publisher}{ACM}, \bibinfo{address}{New York, NY, USA}, \bibinfo{pages}{1850--1863}.
\newblock
\showISBNx{9781450343350}
\urldef\tempurl%
\url{https://doi.org/10.1145/2998181.2998186}
\showDOI{\tempurl}


\bibitem[Keseru and Macharis(2018)]%
        {Keseru2018-bj}
\bibfield{author}{\bibinfo{person}{Imre Keseru} {and} \bibinfo{person}{Cathy Macharis}.} \bibinfo{year}{2018}\natexlab{}.
\newblock \showarticletitle{Travel-based multitasking: review of the empirical evidence}.
\newblock \bibinfo{journal}{\emph{Transp. Rev.}} \bibinfo{volume}{38}, \bibinfo{number}{2} (\bibinfo{date}{March} \bibinfo{year}{2018}), \bibinfo{pages}{162--183}.
\newblock


\bibitem[Kim et~al\mbox{.}(2014)]%
        {Kim2014-fk}
\bibfield{author}{\bibinfo{person}{Sung~Doo Kim}, \bibinfo{person}{Daniele~A Bologna}, \bibinfo{person}{Stacie Furst-Holloway}, \bibinfo{person}{Elaine~C Hollensbe}, \bibinfo{person}{Suzanne~S Masterson}, {and} \bibinfo{person}{Therese Sprinkle}.} \bibinfo{year}{2014}\natexlab{}.
\newblock \showarticletitle{``Taking a break via technology? Triggers, nature, and effects of ''``online''`` work breaks''}.
\newblock \bibinfo{journal}{\emph{Acad. Manag. Proc.}} \bibinfo{volume}{2014}, \bibinfo{number}{1} (\bibinfo{date}{Jan.} \bibinfo{year}{2014}), \bibinfo{pages}{11891}.
\newblock


\bibitem[Lane et~al\mbox{.}(2007)]%
        {Lane2007-jn}
\bibfield{author}{\bibinfo{person}{James~D Lane}, \bibinfo{person}{Jon~E Seskevich}, {and} \bibinfo{person}{Carl~F Pieper}.} \bibinfo{year}{2007}\natexlab{}.
\newblock \showarticletitle{Brief meditation training can improve perceived stress and negative mood}.
\newblock \bibinfo{journal}{\emph{Altern. Ther. Health Med.}} \bibinfo{volume}{13}, \bibinfo{number}{1} (\bibinfo{year}{2007}), \bibinfo{pages}{38--44}.
\newblock
\showISSN{1078-6791}
\urldef\tempurl%
\url{https://www.ncbi.nlm.nih.gov/pubmed/17283740}
\showURL{%
\tempurl}


\bibitem[Larson et~al\mbox{.}(2001)]%
        {Larson2001-pa}
\bibfield{author}{\bibinfo{person}{R~W Larson}, \bibinfo{person}{M~H Richards}, \bibinfo{person}{B Sims}, {and} \bibinfo{person}{J Dworkin}.} \bibinfo{year}{2001}\natexlab{}.
\newblock \showarticletitle{How urban African American young adolescents spend their time: time budgets for locations, activities, and companionship}.
\newblock \bibinfo{journal}{\emph{Am. J. Community Psychol.}} \bibinfo{volume}{29}, \bibinfo{number}{4} (\bibinfo{date}{Aug.} \bibinfo{year}{2001}), \bibinfo{pages}{565--597}.
\newblock
\showISSN{0091-0562, 1573-2770}
\urldef\tempurl%
\url{https://doi.org/10.1023/A:1010422017731}
\showDOI{\tempurl}


\bibitem[Lee et~al\mbox{.}(2023)]%
        {Lee2023-xo}
\bibfield{author}{\bibinfo{person}{Hsinju Lee}, \bibinfo{person}{Fang-Hsin Hsu}, \bibinfo{person}{Wei-Ko Li}, \bibinfo{person}{Jie Tsai}, \bibinfo{person}{Ying-Yu Chen}, {and} \bibinfo{person}{Yung-Ju Chang}.} \bibinfo{year}{2023}\natexlab{}.
\newblock \showarticletitle{Get distracted or missed the stop? Investigating public transit passengers' travel-based multitasking behaviors, motives, and challenges}. In \bibinfo{booktitle}{\emph{Proceedings of the 2023 {CHI} Conference on Human Factors in Computing Systems}} (Hamburg Germany) \emph{(\bibinfo{series}{CHI '23}, \bibinfo{number}{Article 751})}. \bibinfo{publisher}{ACM}, \bibinfo{address}{New York, NY, USA}, \bibinfo{pages}{1--14}.
\newblock
\showISBNx{9781450394215}
\urldef\tempurl%
\url{https://doi.org/10.1145/3544548.3581391}
\showDOI{\tempurl}


\bibitem[Lejeune and Wearden(2009)]%
        {Lejeune2009-eu}
\bibfield{author}{\bibinfo{person}{Helga Lejeune} {and} \bibinfo{person}{J~H Wearden}.} \bibinfo{year}{2009}\natexlab{}.
\newblock \showarticletitle{{Vierordt'sThe} Experimental Study of the Time Sense(1868) and its legacy}.
\newblock \bibinfo{journal}{\emph{Eur. J. Cogn. Psychol.}} \bibinfo{volume}{21}, \bibinfo{number}{6} (\bibinfo{date}{Sept.} \bibinfo{year}{2009}), \bibinfo{pages}{941--960}.
\newblock
\showISSN{0954-1446, 1464-0635}
\urldef\tempurl%
\url{https://doi.org/10.1080/09541440802453006}
\showDOI{\tempurl}


\bibitem[Leshed and Sengers(2011)]%
        {Leshed2011-zp}
\bibfield{author}{\bibinfo{person}{Gilly Leshed} {and} \bibinfo{person}{Phoebe Sengers}.} \bibinfo{year}{2011}\natexlab{}.
\newblock \showarticletitle{{I} lie to myself that i have freedom in my own schedule: productivity tools and experiences of busyness}. In \bibinfo{booktitle}{\emph{Proceedings of the {SIGCHI} Conference on Human Factors in Computing Systems}} (Vancouver BC Canada) \emph{(\bibinfo{series}{CHI '11})}. \bibinfo{publisher}{ACM}, \bibinfo{address}{New York, NY, USA}, \bibinfo{pages}{905--914}.
\newblock
\showISBNx{9781450302289}
\urldef\tempurl%
\url{https://doi.org/10.1145/1978942.1979077}
\showDOI{\tempurl}


\bibitem[Lindley(2015)]%
        {Lindley2015-hv}
\bibfield{author}{\bibinfo{person}{Si{\^a}n~E Lindley}.} \bibinfo{year}{2015}\natexlab{}.
\newblock \showarticletitle{Making Time}. In \bibinfo{booktitle}{\emph{Proceedings of the 18th {ACM} Conference on Computer Supported Cooperative Work \& Social Computing}} (Vancouver BC Canada) \emph{(\bibinfo{series}{CSCW '15})}. \bibinfo{publisher}{ACM}, \bibinfo{address}{New York, NY, USA}, \bibinfo{pages}{1442--1452}.
\newblock
\showISBNx{9781450329224}
\urldef\tempurl%
\url{https://doi.org/10.1145/2675133.2675157}
\showDOI{\tempurl}


\bibitem[Lukoff et~al\mbox{.}(2023)]%
        {Lukoff2023-lt}
\bibfield{author}{\bibinfo{person}{Kai Lukoff}, \bibinfo{person}{Ulrik Lyngs}, \bibinfo{person}{Karina Shirokova}, \bibinfo{person}{Raveena Rao}, \bibinfo{person}{Larry Tian}, \bibinfo{person}{Himanshu Zade}, \bibinfo{person}{Sean~A Munson}, {and} \bibinfo{person}{Alexis Hiniker}.} \bibinfo{year}{2023}\natexlab{}.
\newblock \showarticletitle{{SwitchTube}: A {Proof-of-Concept} System Introducing ``Adaptable Commitment Interfaces'' as a Tool for Digital Wellbeing}. In \bibinfo{booktitle}{\emph{Proceedings of the 2023 {CHI} Conference on Human Factors in Computing Systems}} (Hamburg, Germany) \emph{(\bibinfo{series}{CHI '23}, \bibinfo{number}{Article 197})}. \bibinfo{publisher}{Association for Computing Machinery}, \bibinfo{address}{New York, NY, USA}, \bibinfo{pages}{1--22}.
\newblock
\showISBNx{9781450394215}
\urldef\tempurl%
\url{https://doi.org/10.1145/3544548.3580703}
\showDOI{\tempurl}


\bibitem[Lukoff et~al\mbox{.}(2018)]%
        {Lukoff2018-ws}
\bibfield{author}{\bibinfo{person}{Kai Lukoff}, \bibinfo{person}{Cissy Yu}, \bibinfo{person}{Julie Kientz}, {and} \bibinfo{person}{Alexis Hiniker}.} \bibinfo{year}{2018}\natexlab{}.
\newblock \showarticletitle{What makes smartphone use meaningful or meaningless?}
\newblock \bibinfo{journal}{\emph{Proc. ACM Interact. Mob. Wearable Ubiquitous Technol.}} \bibinfo{volume}{2}, \bibinfo{number}{1} (\bibinfo{date}{March} \bibinfo{year}{2018}), \bibinfo{pages}{1--26}.
\newblock
\showISSN{2474-9567}
\urldef\tempurl%
\url{https://doi.org/10.1145/3191754}
\showDOI{\tempurl}


\bibitem[Luo et~al\mbox{.}(2022)]%
        {Luo2022-qm}
\bibfield{author}{\bibinfo{person}{Yuhan Luo}, \bibinfo{person}{Bongshin Lee}, \bibinfo{person}{Young-Ho Kim}, {and} \bibinfo{person}{Eun~Kyoung Choe}.} \bibinfo{year}{2022}\natexlab{}.
\newblock \showarticletitle{{NoteWordy}: Investigating touch and speech input on smartphones for personal data capture}.
\newblock \bibinfo{journal}{\emph{Proc. ACM Hum. Comput. Interact.}} \bibinfo{volume}{6}, \bibinfo{number}{ISS} (\bibinfo{date}{Nov.} \bibinfo{year}{2022}), \bibinfo{pages}{568--591}.
\newblock
\showISSN{2573-0142}
\urldef\tempurl%
\url{https://doi.org/10.1145/3567734}
\showDOI{\tempurl}


\bibitem[Lyons and Urry(2005)]%
        {Lyons2005-rh}
\bibfield{author}{\bibinfo{person}{Glenn Lyons} {and} \bibinfo{person}{John Urry}.} \bibinfo{year}{2005}\natexlab{}.
\newblock \showarticletitle{Travel time use in the information age}.
\newblock \bibinfo{journal}{\emph{Transp. Res. Part A Policy Pract.}} \bibinfo{volume}{39}, \bibinfo{number}{2-3} (\bibinfo{date}{Feb.} \bibinfo{year}{2005}), \bibinfo{pages}{257--276}.
\newblock


\bibitem[Mark et~al\mbox{.}(2014)]%
        {Mark2014-yh}
\bibfield{author}{\bibinfo{person}{Gloria Mark}, \bibinfo{person}{Shamsi Iqbal}, \bibinfo{person}{Mary Czerwinski}, {and} \bibinfo{person}{Paul Johns}.} \bibinfo{year}{2014}\natexlab{}.
\newblock \showarticletitle{Capturing the mood: facebook and face-to-face encounters in the workplace}. In \bibinfo{booktitle}{\emph{Proceedings of the 17th ACM conference on Computer supported cooperative work \& social computing}}. \bibinfo{publisher}{ACM}, \bibinfo{address}{New York, NY, USA}, \bibinfo{pages}{1082--1094}.
\newblock


\bibitem[Mark et~al\mbox{.}(2015)]%
        {Mark2015-ad}
\bibfield{author}{\bibinfo{person}{Gloria Mark}, \bibinfo{person}{Shamsi Iqbal}, \bibinfo{person}{Mary Czerwinski}, {and} \bibinfo{person}{Paul Johns}.} \bibinfo{year}{2015}\natexlab{}.
\newblock \showarticletitle{Focused, aroused, but so distractible: Temporal perspectives on multitasking and communications}. In \bibinfo{booktitle}{\emph{Proceedings of the 18th ACM Conference on Computer Supported Cooperative Work \& Social Computing}} \emph{(\bibinfo{series}{CSCW '15})}. \bibinfo{publisher}{ACM}, \bibinfo{address}{New York, NY, USA}, \bibinfo{pages}{903--916}.
\newblock


\bibitem[Michie et~al\mbox{.}(2011)]%
        {Michie2011-nl}
\bibfield{author}{\bibinfo{person}{Susan Michie}, \bibinfo{person}{Maartje~M van Stralen}, {and} \bibinfo{person}{Robert West}.} \bibinfo{year}{2011}\natexlab{}.
\newblock \showarticletitle{The behaviour change wheel: a new method for characterising and designing behaviour change interventions}.
\newblock \bibinfo{journal}{\emph{Implement. Sci.}} \bibinfo{volume}{6}, \bibinfo{number}{1} (\bibinfo{date}{April} \bibinfo{year}{2011}), \bibinfo{pages}{42}.
\newblock
\showISSN{1748-5908}
\urldef\tempurl%
\url{https://doi.org/10.1186/1748-5908-6-42}
\showDOI{\tempurl}


\bibitem[Ochs and Sauer(2022)]%
        {Ochs2022-fw}
\bibfield{author}{\bibinfo{person}{Carli Ochs} {and} \bibinfo{person}{Juergen Sauer}.} \bibinfo{year}{2022}\natexlab{}.
\newblock \showarticletitle{Disturbing aspects of smartphone usage: a qualitative analysis}.
\newblock \bibinfo{journal}{\emph{Behav. Inf. Technol.}} \bibinfo{volume}{42}, \bibinfo{number}{14} (\bibinfo{date}{Oct.} \bibinfo{year}{2022}), \bibinfo{pages}{2504--2519}.
\newblock


\bibitem[Odom et~al\mbox{.}(2018)]%
        {Odom2018-dv}
\bibfield{author}{\bibinfo{person}{William Odom}, \bibinfo{person}{Si{\^a}n Lindley}, \bibinfo{person}{Larissa Pschetz}, \bibinfo{person}{Vasiliki Tsaknaki}, \bibinfo{person}{Anna Vallg{\aa}rda}, \bibinfo{person}{Mikael Wiberg}, {and} \bibinfo{person}{Daisy Yoo}.} \bibinfo{year}{2018}\natexlab{}.
\newblock \showarticletitle{Time, temporality, and slowness: Future directions for design research}. In \bibinfo{booktitle}{\emph{Proceedings of the 2018 {ACM} Conference Companion Publication on Designing Interactive Systems}} (Hong Kong China) \emph{(\bibinfo{series}{DIS '18 Companion})}. \bibinfo{publisher}{ACM}, \bibinfo{address}{New York, NY, USA}, \bibinfo{pages}{383--386}.
\newblock
\showISBNx{9781450356312}
\urldef\tempurl%
\url{https://doi.org/10.1145/3197391.3197392}
\showDOI{\tempurl}


\bibitem[Odom et~al\mbox{.}(2022)]%
        {Odom2022-aj}
\bibfield{author}{\bibinfo{person}{William Odom}, \bibinfo{person}{Erik Stolterman}, {and} \bibinfo{person}{Amy Yo~Sue Chen}.} \bibinfo{year}{2022}\natexlab{}.
\newblock \showarticletitle{Extending a theory of slow technology for design through artifact analysis}.
\newblock \bibinfo{journal}{\emph{Hum.-Comput. Interact.}} \bibinfo{volume}{37}, \bibinfo{number}{2} (\bibinfo{date}{March} \bibinfo{year}{2022}), \bibinfo{pages}{150--179}.
\newblock
\showISSN{0737-0024, 1532-7051}
\urldef\tempurl%
\url{https://doi.org/10.1080/07370024.2021.1913416}
\showDOI{\tempurl}


\bibitem[of~Labor~Statistics(2021)]%
        {ATUS2021-mw}
\bibfield{author}{\bibinfo{person}{United States.~Bureau of Labor~Statistics}.} \bibinfo{year}{2021}\natexlab{}.
\newblock \bibinfo{title}{Data sources: Handbook of Methods}.
\newblock \bibinfo{howpublished}{\url{https://www.bls.gov/opub/hom/atus/data.htm}}.
\newblock
\newblock
\shownote{Accessed: 2023-12-10}.


\bibitem[of~Labor~Statistics(2022a)]%
        {ATUS2022-bc}
\bibfield{author}{\bibinfo{person}{United States.~Bureau of Labor~Statistics}.} \bibinfo{year}{2022}\natexlab{a}.
\newblock \bibinfo{title}{{ATUS} home}.
\newblock \bibinfo{howpublished}{\url{https://www.bls.gov/tus/}}.
\newblock
\newblock
\shownote{Accessed: 2023-12-10}.


\bibitem[of~Labor~Statistics(2022b)]%
        {ATUS2022-qc}
\bibfield{author}{\bibinfo{person}{United States.~Bureau of Labor~Statistics}.} \bibinfo{year}{2022}\natexlab{b}.
\newblock \bibinfo{title}{Table 3. Time spent in primary activities for the civilian population by age, sex, race, Hispanic or Latino ethnicity, marital status, and educational attainment, 2022 annual averages - 2022 {A01} Results}.
\newblock
\newblock
\urldef\tempurl%
\url{https://www.bls.gov/news.release/atus.t03.htm}
\showURL{%
\tempurl}
\newblock
\shownote{Accessed: 2023-9-6}.


\bibitem[of~Labor~Statistics(2023)]%
        {ATUS2023-jr}
\bibfield{author}{\bibinfo{person}{United States.~Bureau of Labor~Statistics}.} \bibinfo{year}{2023}\natexlab{}.
\newblock \bibinfo{title}{{ATUS} Activity Coding Lexicons and Coding Manuals}.
\newblock \bibinfo{howpublished}{\url{https://www.bls.gov/tus/lexicons.htm}}.
\newblock
\newblock
\shownote{Accessed: 2023-12-10}.


\bibitem[Oulasvirta et~al\mbox{.}(2012)]%
        {Oulasvirta2012-en}
\bibfield{author}{\bibinfo{person}{Antti Oulasvirta}, \bibinfo{person}{Tye Rattenbury}, \bibinfo{person}{Lingyi Ma}, {and} \bibinfo{person}{Eeva Raita}.} \bibinfo{year}{2012}\natexlab{}.
\newblock \showarticletitle{Habits make smartphone use more pervasive}.
\newblock \bibinfo{journal}{\emph{Pers. Ubiquit. Comput.}} \bibinfo{volume}{16}, \bibinfo{number}{1} (\bibinfo{date}{Jan.} \bibinfo{year}{2012}), \bibinfo{pages}{105--114}.
\newblock
\showISSN{0949-2054}
\urldef\tempurl%
\url{https://doi.org/10.1007/s00779-011-0412-2}
\showDOI{\tempurl}


\bibitem[Ozmen~Garibay et~al\mbox{.}(2023)]%
        {Ozmen_Garibay2023-ct}
\bibfield{author}{\bibinfo{person}{Ozlem Ozmen~Garibay}, \bibinfo{person}{Brent Winslow}, \bibinfo{person}{Salvatore Andolina}, \bibinfo{person}{Margherita Antona}, \bibinfo{person}{Anja Bodenschatz}, \bibinfo{person}{Constantinos Coursaris}, \bibinfo{person}{Gregory Falco}, \bibinfo{person}{Stephen~M Fiore}, \bibinfo{person}{Ivan Garibay}, \bibinfo{person}{Keri Grieman}, \bibinfo{person}{John~C Havens}, \bibinfo{person}{Marina Jirotka}, \bibinfo{person}{Hernisa Kacorri}, \bibinfo{person}{Waldemar Karwowski}, \bibinfo{person}{Joe Kider}, \bibinfo{person}{Joseph Konstan}, \bibinfo{person}{Sean Koon}, \bibinfo{person}{Monica Lopez-Gonzalez}, \bibinfo{person}{Iliana Maifeld-Carucci}, \bibinfo{person}{Sean McGregor}, \bibinfo{person}{Gavriel Salvendy}, \bibinfo{person}{Ben Shneiderman}, \bibinfo{person}{Constantine Stephanidis}, \bibinfo{person}{Christina Strobel}, \bibinfo{person}{Carolyn Ten~Holter}, {and} \bibinfo{person}{Wei Xu}.} \bibinfo{year}{2023}\natexlab{}.
\newblock \showarticletitle{Six human-centered artificial intelligence grand challenges}.
\newblock \bibinfo{journal}{\emph{Int. J. Hum. Comput. Interact.}} \bibinfo{volume}{39}, \bibinfo{number}{3} (\bibinfo{date}{Jan.} \bibinfo{year}{2023}), \bibinfo{pages}{1--47}.
\newblock
\showISSN{1044-7318, 1532-7590}
\urldef\tempurl%
\url{https://doi.org/10.1080/10447318.2022.2153320}
\showDOI{\tempurl}


\bibitem[Rahm-Sk{\aa}geby and Rahm(2022)]%
        {Rahm-Skageby2022-iv}
\bibfield{author}{\bibinfo{person}{J{\"o}rgen Rahm-Sk{\aa}geby} {and} \bibinfo{person}{Lina Rahm}.} \bibinfo{year}{2022}\natexlab{}.
\newblock \showarticletitle{{HCI} and deep time: toward deep time design thinking}.
\newblock \bibinfo{journal}{\emph{Hum.-Comput. Interact.}} \bibinfo{volume}{37}, \bibinfo{number}{1} (\bibinfo{date}{Jan.} \bibinfo{year}{2022}), \bibinfo{pages}{15--28}.
\newblock
\showISSN{0737-0024, 1532-7051}
\urldef\tempurl%
\url{https://doi.org/10.1080/07370024.2021.1902328}
\showDOI{\tempurl}


\bibitem[Rattenbury et~al\mbox{.}(2008)]%
        {Rattenbury2008-gz}
\bibfield{author}{\bibinfo{person}{Tye Rattenbury}, \bibinfo{person}{Dawn Nafus}, {and} \bibinfo{person}{Ken Anderson}.} \bibinfo{year}{2008}\natexlab{}.
\newblock \showarticletitle{Plastic: a metaphor for integrated technologies}. In \bibinfo{booktitle}{\emph{Proceedings of the 10th international conference on Ubiquitous computing}} (Seoul Korea) \emph{(\bibinfo{series}{UbiComp '08})}. \bibinfo{publisher}{ACM}, \bibinfo{address}{New York, NY, USA}, \bibinfo{pages}{232--241}.
\newblock
\showISBNx{9781605581361}
\urldef\tempurl%
\url{https://doi.org/10.1145/1409635.1409667}
\showDOI{\tempurl}


\bibitem[Ruth~Eikhof(2007)]%
        {Ruth_Eikhof2007-pl}
\bibfield{author}{\bibinfo{person}{Doris Ruth~Eikhof}.} \bibinfo{year}{2007}\natexlab{}.
\newblock \showarticletitle{Introduction: What work? What life? What balance?}
\newblock \bibinfo{journal}{\emph{Empl. Relat.}} \bibinfo{volume}{29}, \bibinfo{number}{4} (\bibinfo{date}{July} \bibinfo{year}{2007}), \bibinfo{pages}{325--333}.
\newblock
\showISSN{0142-5455, 1758-7069}
\urldef\tempurl%
\url{https://doi.org/10.1108/er.2007.01929daa.001}
\showDOI{\tempurl}


\bibitem[Schor(2008)]%
        {Schor2008-mq}
\bibfield{author}{\bibinfo{person}{J Schor}.} \bibinfo{year}{2008}\natexlab{}.
\newblock \bibinfo{booktitle}{\emph{The overworked American: The unexpected decline of leisure}}.
\newblock \bibinfo{publisher}{Basic Books}, \bibinfo{address}{London, England}.
\newblock


\bibitem[Shudong and Higgins(2005)]%
        {Shudong2005-qr}
\bibfield{author}{\bibinfo{person}{Wang Shudong} {and} \bibinfo{person}{M Higgins}.} \bibinfo{year}{2005}\natexlab{}.
\newblock \showarticletitle{Limitations of mobile phone learning}. In \bibinfo{booktitle}{\emph{{IEEE} International Workshop on Wireless and Mobile Technologies in Education ({WMTE'05})}} (Tokushima, Japan). \bibinfo{publisher}{IEEE}, \bibinfo{address}{New York, NY, USA}, \bibinfo{pages}{3 pp.--181}.
\newblock
\showISBNx{9780769523859}
\urldef\tempurl%
\url{https://doi.org/10.1109/WMTE.2005.43}
\showDOI{\tempurl}


\bibitem[Skatova et~al\mbox{.}(2016)]%
        {Skatova2016-lx}
\bibfield{author}{\bibinfo{person}{Anya Skatova}, \bibinfo{person}{Ben Bedwell}, \bibinfo{person}{Victoria Shipp}, \bibinfo{person}{Yitong Huang}, \bibinfo{person}{Alexandra Young}, \bibinfo{person}{Tom Rodden}, {and} \bibinfo{person}{Emma Bertenshaw}.} \bibinfo{year}{2016}\natexlab{}.
\newblock \showarticletitle{The Role of {ICT} in Office Work Breaks}. In \bibinfo{booktitle}{\emph{Proceedings of the 2016 {CHI} Conference on Human Factors in Computing Systems}} (San Jose, California, USA) \emph{(\bibinfo{series}{CHI '16})}. \bibinfo{publisher}{Association for Computing Machinery}, \bibinfo{address}{New York, NY, USA}, \bibinfo{pages}{3049--3060}.
\newblock
\showISBNx{9781450333627}
\urldef\tempurl%
\url{https://doi.org/10.1145/2858036.2858443}
\showDOI{\tempurl}


\bibitem[Skimina et~al\mbox{.}(2020)]%
        {Skimina2020-kx}
\bibfield{author}{\bibinfo{person}{Ewa Skimina}, \bibinfo{person}{Dominika Kara{\'s}}, \bibinfo{person}{Ewa Topolewska-Siedzik}, \bibinfo{person}{Maria K{\l}ym-Guba}, \bibinfo{person}{Klaudia Ponikiewska}, \bibinfo{person}{Rados{\l}aw Rogoza}, \bibinfo{person}{Eldad Davidov}, {and} \bibinfo{person}{Jan Cieciuch}.} \bibinfo{year}{2020}\natexlab{}.
\newblock \showarticletitle{A categorization of behaviors reported in experience sampling studies}.
\newblock \bibinfo{journal}{\emph{Soc. Psychol. Bull.}} \bibinfo{volume}{15}, \bibinfo{number}{2} (\bibinfo{date}{July} \bibinfo{year}{2020}), \bibinfo{pages}{1--33}.
\newblock
\showISSN{1896-1800, 2569-653X}
\urldef\tempurl%
\url{https://doi.org/10.32872/spb.3029}
\showDOI{\tempurl}


\bibitem[Stephanidis et~al\mbox{.}(2019)]%
        {Stephanidis2019-hl}
\bibfield{author}{\bibinfo{person}{Constantine Stephanidis}, \bibinfo{person}{Gavriel Salvendy}, \bibinfo{person}{Margherita Antona}, \bibinfo{person}{Jessie Y~C Chen}, \bibinfo{person}{Jianming Dong}, \bibinfo{person}{Vincent~G Duffy}, \bibinfo{person}{Xiaowen Fang}, \bibinfo{person}{Cali Fidopiastis}, \bibinfo{person}{Gino Fragomeni}, \bibinfo{person}{Limin~Paul Fu}, \bibinfo{person}{Yinni Guo}, \bibinfo{person}{Don Harris}, \bibinfo{person}{Andri Ioannou}, \bibinfo{person}{Kyeong-Ah~(kate) Jeong}, \bibinfo{person}{Shin'ichi Konomi}, \bibinfo{person}{Heidi Kr{\"o}mker}, \bibinfo{person}{Masaaki Kurosu}, \bibinfo{person}{James~R Lewis}, \bibinfo{person}{Aaron Marcus}, \bibinfo{person}{Gabriele Meiselwitz}, \bibinfo{person}{Abbas Moallem}, \bibinfo{person}{Hirohiko Mori}, \bibinfo{person}{Fiona Fui-Hoon~Nah}, \bibinfo{person}{Stavroula Ntoa}, \bibinfo{person}{Pei-Luen~Patrick Rau}, \bibinfo{person}{Dylan Schmorrow}, \bibinfo{person}{Keng Siau}, \bibinfo{person}{Norbert Streitz}, \bibinfo{person}{Wentao Wang},
  \bibinfo{person}{Sakae Yamamoto}, \bibinfo{person}{Panayiotis Zaphiris}, {and} \bibinfo{person}{Jia Zhou}.} \bibinfo{year}{2019}\natexlab{}.
\newblock \showarticletitle{Seven {HCI} grand challenges}.
\newblock \bibinfo{journal}{\emph{Int. J. Hum. Comput. Interact.}} \bibinfo{volume}{35}, \bibinfo{number}{14} (\bibinfo{date}{Aug.} \bibinfo{year}{2019}), \bibinfo{pages}{1229--1269}.
\newblock
\showISSN{1044-7318, 1532-7590}
\urldef\tempurl%
\url{https://doi.org/10.1080/10447318.2019.1619259}
\showDOI{\tempurl}


\bibitem[Steup et~al\mbox{.}(2022)]%
        {Steup2022-gx}
\bibfield{author}{\bibinfo{person}{Rosemary Steup}, \bibinfo{person}{Paige White}, \bibinfo{person}{Lynn Dombrowski}, {and} \bibinfo{person}{Norman~Makoto Su}.} \bibinfo{year}{2022}\natexlab{}.
\newblock \showarticletitle{A reasonable life: Rhythmic attunement and sustainable work at the intersection of farming and knowledge work}.
\newblock \bibinfo{journal}{\emph{Proc. ACM Hum. Comput. Interact.}} \bibinfo{volume}{6}, \bibinfo{number}{CSCW2} (\bibinfo{date}{Nov.} \bibinfo{year}{2022}), \bibinfo{pages}{1--22}.
\newblock
\showISSN{2573-0142}
\urldef\tempurl%
\url{https://doi.org/10.1145/3555577}
\showDOI{\tempurl}


\bibitem[Su et~al\mbox{.}(2022)]%
        {Su2022-nm}
\bibfield{author}{\bibinfo{person}{Sizhen Su}, \bibinfo{person}{Le Shi}, \bibinfo{person}{Yongbo Zheng}, \bibinfo{person}{Yankun Sun}, \bibinfo{person}{Xiaolin Huang}, \bibinfo{person}{Anyi Zhang}, \bibinfo{person}{Jianyu Que}, \bibinfo{person}{Xinyu Sun}, \bibinfo{person}{Jie Shi}, \bibinfo{person}{Yanping Bao}, \bibinfo{person}{Jiahui Deng}, {and} \bibinfo{person}{Lin Lu}.} \bibinfo{year}{2022}\natexlab{}.
\newblock \showarticletitle{Leisure activities and the risk of dementia: A systematic review and meta-analysis: A systematic review and meta-analysis}.
\newblock \bibinfo{journal}{\emph{Neurology}} \bibinfo{volume}{99}, \bibinfo{number}{15} (\bibinfo{date}{Aug.} \bibinfo{year}{2022}), \bibinfo{pages}{e1651--e1663}.
\newblock
\showISSN{0028-3878, 1526-632X}
\urldef\tempurl%
\url{https://doi.org/10.1212/WNL.0000000000200929}
\showDOI{\tempurl}


\bibitem[Swearngin et~al\mbox{.}(2021)]%
        {Swearngin2021-ys}
\bibfield{author}{\bibinfo{person}{Amanda Swearngin}, \bibinfo{person}{Shamsi Iqbal}, \bibinfo{person}{Victor Poznanski}, \bibinfo{person}{Mark Encarnaci{\'o}n}, \bibinfo{person}{Paul~N Bennett}, {and} \bibinfo{person}{Jaime Teevan}.} \bibinfo{year}{2021}\natexlab{}.
\newblock \showarticletitle{Scraps: Enabling Mobile Capture, Contextualization, and Use of Document Resources}. In \bibinfo{booktitle}{\emph{Proceedings of the 2021 {CHI} Conference on Human Factors in Computing Systems}} (Yokohama, Japan) \emph{(\bibinfo{series}{CHI '21}, \bibinfo{number}{Article 641})}. \bibinfo{publisher}{Association for Computing Machinery}, \bibinfo{address}{New York, NY, USA}, \bibinfo{pages}{1--14}.
\newblock
\showISBNx{9781450380966}
\urldef\tempurl%
\url{https://doi.org/10.1145/3411764.3445185}
\showDOI{\tempurl}


\bibitem[Tandon et~al\mbox{.}(2022)]%
        {Tandon2022-ki}
\bibfield{author}{\bibinfo{person}{Anushree Tandon}, \bibinfo{person}{Amandeep Dhir}, \bibinfo{person}{Shalini Talwar}, \bibinfo{person}{Puneet Kaur}, {and} \bibinfo{person}{Matti Mäntymäki}.} \bibinfo{year}{2022}\natexlab{}.
\newblock \showarticletitle{Social media induced fear of missing out ({FoMO}) and phubbing: Behavioural, relational and psychological outcomes}.
\newblock \bibinfo{journal}{\emph{Technol. Forecast. Soc. Change}} \bibinfo{volume}{174}, \bibinfo{number}{121149} (\bibinfo{date}{Jan.} \bibinfo{year}{2022}), \bibinfo{pages}{121149}.
\newblock


\bibitem[Teevan(2016)]%
        {Teevan2016-ed}
\bibfield{author}{\bibinfo{person}{Jaime Teevan}.} \bibinfo{year}{2016}\natexlab{}.
\newblock \showarticletitle{The future of microwork}.
\newblock \bibinfo{journal}{\emph{Crossroads}} \bibinfo{volume}{23}, \bibinfo{number}{2} (\bibinfo{date}{Dec.} \bibinfo{year}{2016}), \bibinfo{pages}{26--29}.
\newblock
\showISSN{1528-4972, 1528-4980}
\urldef\tempurl%
\url{https://doi.org/10.1145/3019600}
\showDOI{\tempurl}


\bibitem[Terzimehi{\'c} and Aragon-Hahner(2022)]%
        {Terzimehic2022-si}
\bibfield{author}{\bibinfo{person}{Na{\dj}a Terzimehi{\'c}} {and} \bibinfo{person}{Sarah Aragon-Hahner}.} \bibinfo{year}{2022}\natexlab{}.
\newblock \showarticletitle{{I} wish {I} had: Desired real-world activities instead of regretful smartphone use}. In \bibinfo{booktitle}{\emph{Proceedings of the 21st International Conference on Mobile and Ubiquitous Multimedia}} (Lisbon Portugal) \emph{(\bibinfo{series}{MUM '22})}. \bibinfo{publisher}{ACM}, \bibinfo{address}{New York, NY, USA}, \bibinfo{pages}{47--52}.
\newblock
\showISBNx{9781450398206}
\urldef\tempurl%
\url{https://doi.org/10.1145/3568444.3568465}
\showDOI{\tempurl}


\bibitem[Terzimehic et~al\mbox{.}(2023)]%
        {Terzimehic2023-yi}
\bibfield{author}{\bibinfo{person}{Nada Terzimehic}, \bibinfo{person}{Florian Bemmann}, \bibinfo{person}{Miriam Halsner}, {and} \bibinfo{person}{Sven Mayer}.} \bibinfo{year}{2023}\natexlab{}.
\newblock \showarticletitle{A Mixed-Method Exploration into the Mobile Phone Rabbit Hole}.
\newblock \bibinfo{journal}{\emph{Proc. ACM Hum.-Comput. Interact.}} \bibinfo{volume}{7}, \bibinfo{number}{MHCI} (\bibinfo{date}{Sept.} \bibinfo{year}{2023}), \bibinfo{pages}{1--29}.
\newblock


\bibitem[van Berkel et~al\mbox{.}(2018)]%
        {van-Berkel2018-wr}
\bibfield{author}{\bibinfo{person}{Niels van Berkel}, \bibinfo{person}{Denzil Ferreira}, {and} \bibinfo{person}{Vassilis Kostakos}.} \bibinfo{year}{2018}\natexlab{}.
\newblock \showarticletitle{The Experience Sampling Method on mobile devices}.
\newblock \bibinfo{journal}{\emph{ACM Comput. Surv.}} \bibinfo{volume}{50}, \bibinfo{number}{6} (\bibinfo{date}{Nov.} \bibinfo{year}{2018}), \bibinfo{pages}{1--40}.
\newblock
\showISSN{0360-0300, 1557-7341}
\urldef\tempurl%
\url{https://doi.org/10.1145/3123988}
\showDOI{\tempurl}


\bibitem[Wiberg and Stolterman(2021)]%
        {Wiberg2021-tx}
\bibfield{author}{\bibinfo{person}{Mikael Wiberg} {and} \bibinfo{person}{Erik Stolterman}.} \bibinfo{year}{2021}\natexlab{}.
\newblock \showarticletitle{Time and temporality in {HCI} research}.
\newblock \bibinfo{journal}{\emph{Interact. Comput.}} \bibinfo{volume}{33}, \bibinfo{number}{3} (\bibinfo{date}{May} \bibinfo{year}{2021}), \bibinfo{pages}{250--270}.
\newblock
\showISSN{0953-5438, 1873-7951}
\urldef\tempurl%
\url{https://doi.org/10.1093/iwc/iwab025}
\showDOI{\tempurl}


\bibitem[Williams et~al\mbox{.}(2019)]%
        {Williams2019-ls}
\bibfield{author}{\bibinfo{person}{Alex~C Williams}, \bibinfo{person}{Gloria Mark}, \bibinfo{person}{Kristy Milland}, \bibinfo{person}{Edward Lank}, {and} \bibinfo{person}{Edith Law}.} \bibinfo{year}{2019}\natexlab{}.
\newblock \showarticletitle{The perpetual work life of crowdworkers: How tooling practices increase fragmentation in crowdwork}.
\newblock \bibinfo{journal}{\emph{Proc. ACM Hum. Comput. Interact.}} \bibinfo{volume}{3}, \bibinfo{number}{CSCW} (\bibinfo{date}{Nov.} \bibinfo{year}{2019}), \bibinfo{pages}{1--28}.
\newblock
\showISSN{2573-0142}
\urldef\tempurl%
\url{https://doi.org/10.1145/3359126}
\showDOI{\tempurl}


\bibitem[Zeidan et~al\mbox{.}(2010)]%
        {Zeidan2010-ps}
\bibfield{author}{\bibinfo{person}{Fadel Zeidan}, \bibinfo{person}{Susan~K Johnson}, \bibinfo{person}{Bruce~J Diamond}, \bibinfo{person}{Zhanna David}, {and} \bibinfo{person}{Paula Goolkasian}.} \bibinfo{year}{2010}\natexlab{}.
\newblock \showarticletitle{Mindfulness meditation improves cognition: evidence of brief mental training}.
\newblock \bibinfo{journal}{\emph{Conscious. Cogn.}} \bibinfo{volume}{19}, \bibinfo{number}{2} (\bibinfo{date}{June} \bibinfo{year}{2010}), \bibinfo{pages}{597--605}.
\newblock
\showISSN{1053-8100, 1090-2376}
\urldef\tempurl%
\url{https://doi.org/10.1016/j.concog.2010.03.014}
\showDOI{\tempurl}


\bibitem[Zhang et~al\mbox{.}(2022)]%
        {Zhang2022-rm}
\bibfield{author}{\bibinfo{person}{Mingrui~Ray Zhang}, \bibinfo{person}{Kai Lukoff}, \bibinfo{person}{Raveena Rao}, \bibinfo{person}{Amanda Baughan}, {and} \bibinfo{person}{Alexis Hiniker}.} \bibinfo{year}{2022}\natexlab{}.
\newblock \showarticletitle{Monitoring screen time or redesigning it?: Two approaches to supporting intentional social media use}. In \bibinfo{booktitle}{\emph{{CHI} Conference on Human Factors in Computing Systems}} (New Orleans LA USA) \emph{(\bibinfo{series}{CHI '22}, \bibinfo{number}{Article 60})}. \bibinfo{publisher}{ACM}, \bibinfo{address}{New York, NY, USA}, \bibinfo{pages}{1--19}.
\newblock
\showISBNx{9781450391573}
\urldef\tempurl%
\url{https://doi.org/10.1145/3491102.3517722}
\showDOI{\tempurl}


\end{thebibliography}

\appendix
\section{Distribution of Waiting Time Duration}
\label{appendix:dur}

\begin{figure}[ht]
  \includegraphics[width=0.4\textwidth]{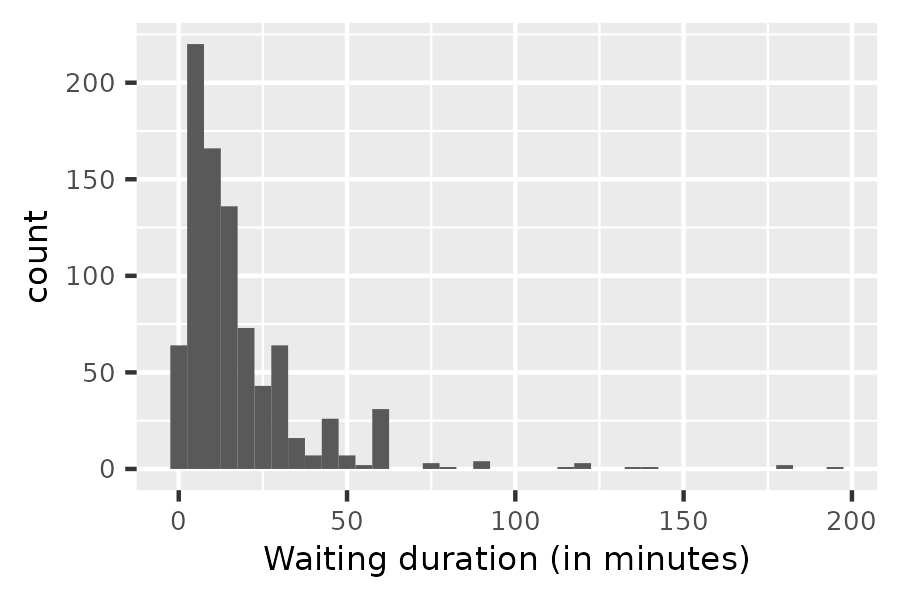}
  \caption{Distribution of the duration (in minutes) of each waiting session.}
  \Description{A histogram with x axis ranging from 0 to 200 and y axis ranging from 0 to 250. Bars cluster around 10.}
  \label{fig:hist_dur}
\end{figure}

\begin{figure}[ht]
  \includegraphics[width=0.4\textwidth]{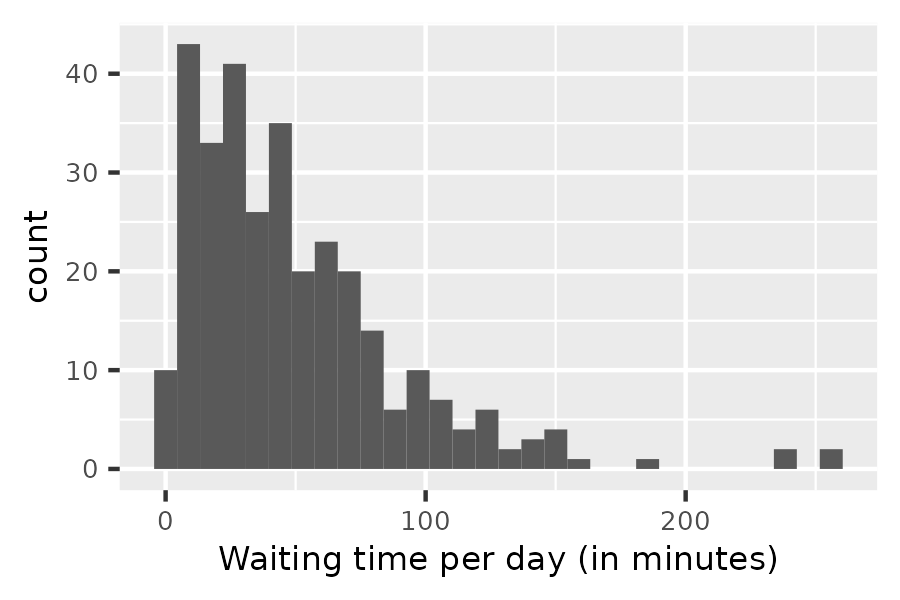}
  \caption{Distribution of the sum duration (in minutes) of waiting sessions in each day.}
  \Description{A histogram with x axis ranging from 0 to 250 and y axis ranging from 0 to 45. Bars cluster between 0 and 100, and 0 to 50 especially, on x axis.}
  \label{fig:hist_dur_day}
\end{figure}

\newpage
\onecolumn
\section{Categories of Waiting Time Activities}
\label{appendix:code}
\begin{table}[h]
\caption{Examples and Percentage (of Duration) of Each Sub-category of Waiting Time Activities}
  \label{table:code}
\begin{tabular}{lllr}
\toprule
Category      & Sub-category                          & Example                                       & Duration\% \\
\midrule
Leisure       & Watching TV, movie, or video  & "watching TV"                                 & 13.4\%     \\
              & Socializing                   & "catch up with friends at table"              & 8.3\%      \\
              & Social media                  & "scrolling facebook"                          & 7.5\%      \\
              & Reading for leisure           & "reading news"                                & 7.0\%      \\
              & Listening to music or podcast & "listening to music, listening to a podcast " & 5.1\%      \\
              & Games                         & "play Disney emoji"                           & 3.7\%      \\
              & Text messages                 & "checking messages"                           & 2.9\%      \\
              & Relaxing or resting           & "relaxing, stretching"                        & 2.3\%      \\
              & Physical activity             & "treadmill, cycling"                          & 2.2\%      \\
              & Phonecall                     & "talking to my friend on the phone"           & 1.5\%      \\
              & Browsing internet             & "on the internet"                             & 1.4\%      \\
              & Using my phone                & "scrolling on phone"                          & 1.0\%      \\
              & Religious practices           & "church services"                             & 0.5\%      \\
              & Time with pets                & "played withy pets outside"                   & 0.5\%      \\
              & Art                           & "played piano"                                & 0.1\%      \\
              & Subtotal                           & \textbackslash{}                              & 57.4\%     \\

\hline
Productive    & Email                         & "checking emails"                             & 11.0\%     \\
              & Work                          & "work"                                        & 10.8\%     \\
              & Study                         & "recitation"                                  & 0.6\%      \\
              & Planning \& Preparing         & "made a list"                                 & 0.2\%      \\
              & Subtotal                           & \textbackslash{}                              & 22.5\%     \\
\hline
Maintenance   & Personal care                 & "shower"                                      & 6.5\%      \\
              & Housework                     & "washing dishes"                              & 5.6\%      \\
              & Food \& drink preparing       & "made a salad"                                & 2.3\%      \\
              & Other maintenance             & "clean my glasses"                            & 1.4\%      \\
              & Shopping                      & "ordered some things online"                  & 1.3\%      \\
              & Subtotal                           & \textbackslash{}                              & 17.1\%     \\
\hline
No purpose    & Doing nothing                 & "nothing, wait"                               & 1.7\%      \\
              & Looking around                & "looking out of the window"                   & 0.7\%      \\
              & Standing or sitting           & "just standing"                               & 0.6\%      \\
              & Subtotal                           & \textbackslash{}                              & 3.0\%      \\
All activities & Grand total                         & \textbackslash{}                              & 100.0\%            \\ 
\bottomrule
\end{tabular}
\end{table}

\clearpage
\section{Waiting Time Activities at Different Locations}
\label{appendix:loc}
\begin{figure*}[h]
%  \centering
  \includegraphics[width=\textwidth]{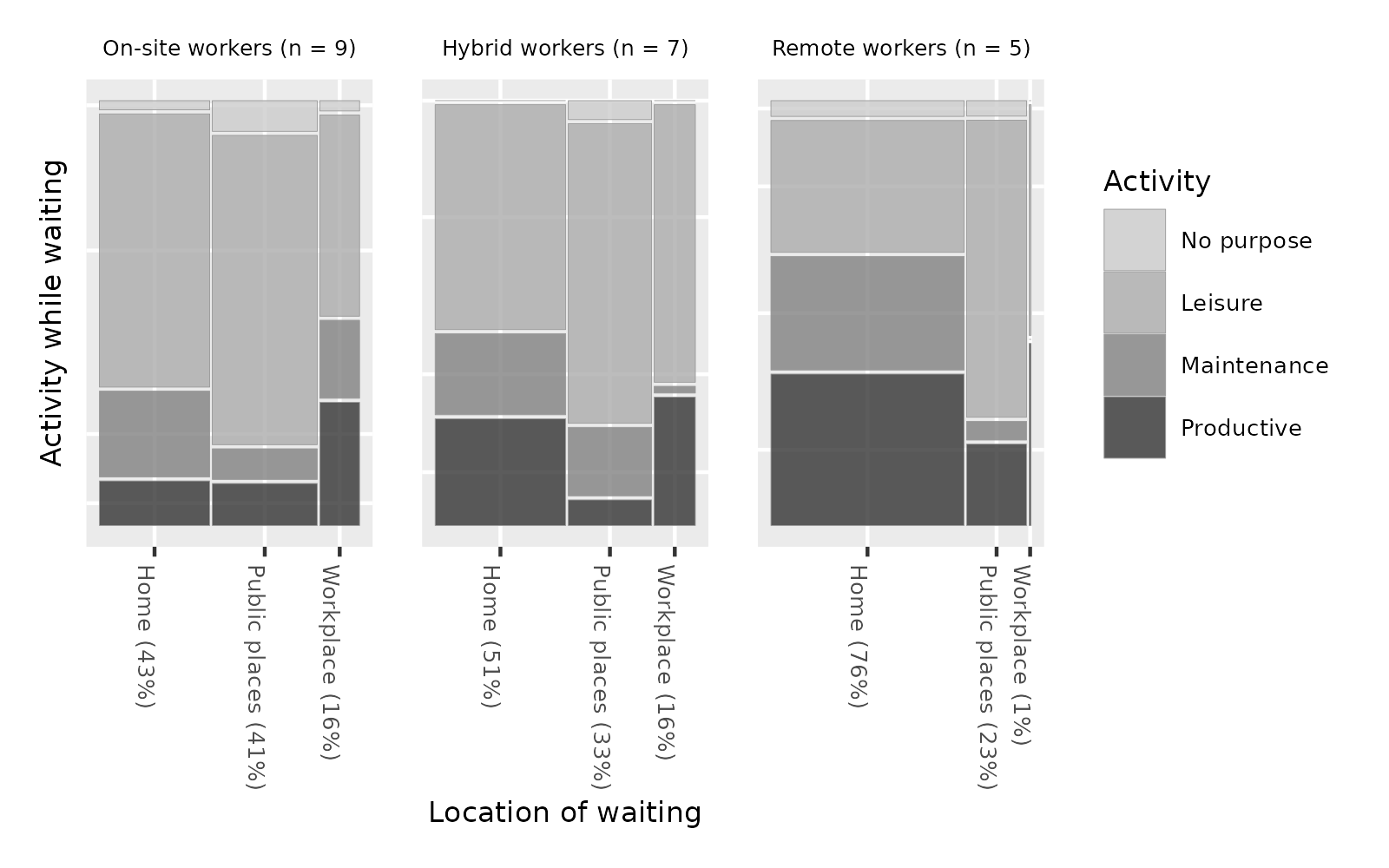}
  \caption{Mosaic plots of waiting time activities at different locations for on-site, hybrid, and remote workers. Areas (and bar widths) indicate time spent in that activity category. The overall patterns, e.g., the association between home-maintenance, public-leisure, and work-productive, are similar across the three subgroups. We also observed that remote workers reported more time waiting at home than on-site and hybrid workers and were more likely to do productive activities while waiting at home. However, the sample size is not large enough to reach robust conclusions or generalize the observation to all working adults.}
  \Description{Three mosaic plots for on-site workers, hybrid workers, and remote workers respectively. Each shows how waiting time is allocated among the four categories, which are leisure, maintenance, productive, and no purpose, at different location of waiting: home, work places, and public places. Comparing across the three locations, leisure is largest at public places, maintenance is largest at home, and productive is largest at work places. The ratio of reported waiting time at home, public, and workplaces was 8:2:0 for remote workers, 5:3:2 for hybrid workers, and 4:4:2 for on-site workers.}
  \label{fig:loc_wp}
\end{figure*}

\clearpage
\section{Logistic Regression Results with log-Transformed Duration}
\label{appendix:regression}
\nopagebreak
\begin{table*}[!hbp]
  \caption{Effects of Situational Factors on Waiting Time Activities with Log-Transformed Duration}
  \label{table:regression_log}
  \begin{tabular}{lrrrrrrrr}
    \toprule
    & \multicolumn{4}{c}{Productive vs. Leisure} & \multicolumn{4}{c}{Maintenance vs. Leisure} \\
    \cmidrule(lr){2-5} \cmidrule(lr){6-9}
    {Variable} & {$B$} & {$SE\ B$} & {$t$} & {$p$} & {$B$} & {$SE\ B$} & {$t$} & {$p$} \\
    \midrule
(Intercept)   & -1.08 & 0.40 & -2.68         & .004                     & -0.97 & 0.39 & -2.47          & .01                      \\
Computer      & 1.54  & 0.25 & \textbf{6.20} & \textbf{\textless{}.001} & -0.08 & 0.24 & -0.34          & .37                      \\
Phone         & 0.19  & 0.50 & 0.37          & .36                      & -0.68 & 0.39 & \textbf{-1.75} & \textbf{.04}             \\
Workplace     & 0.77  & 0.34 & \textbf{2.29} & \textbf{.01}             & 0.37  & 0.46 & 0.79           & .22                      \\
Home          & -0.23 & 0.27 & -0.84         & .20                      & 1.10  & 0.26 & \textbf{4.28}  & \textbf{\textless{}.001} \\
Lunchtime     & 0.72  & 0.26 & \textbf{2.72} & \textbf{.003}            & 0.85  & 0.28 & \textbf{3.02}  & \textbf{.001}            \\
Log-Transformed Duration & 0.13  & 0.11 & 1.16          & .12                      & 0.06  & 0.10 & 0.56           & .29 
\\
    \bottomrule
    \addlinespace
    \multicolumn{9}{l}{\footnotesize *Bold font indicates statistical significance.}
  \end{tabular}
\end{table*}
\end{document}